\documentclass[bibliography,superscriptaddress,pra,twocolumn]{revtex4-2}
 
\usepackage{graphicx,amsmath} 
\usepackage{appendix}
\usepackage{xcolor}
\usepackage{lineno}
\usepackage{soul}

\begin{document}
\title{Towards a micromechanical qubit based on quantized oscillations in superfluid helium}
\author{Priya Sharma$^*$}
\affiliation{School of Mathematics and Physics, University of Surrey, Guildford GU2 7XH, Surrey, UK}
\author{Jens Koch}
\affiliation{Department of Physics and Astronomy, Northwestern University, Evanston IL 60208, USA}
\author{Eran Ginossar}
\affiliation{School of Mathematics and Physics, University of Surrey, Guildford GU2 7XH, Surrey, UK\\$^*$(ps0072\@surrey.ac.uk)}
\date{\today}

\begin{abstract}
{\bf{Abstract:}} Superconducting circuits can exhibit quantized energy levels and long coherence times. Harnessing the anharmonicity offered by Josephson junctions, such circuits have been successfully employed as qubits, quantum limited amplifiers and sensors. Here, we consider superfluidity as the charge-neutral analogue of superconductivity. Both dissipationless mass flow and Josephson tunneling have been demonstrated in superfluid helium. We propose a quantum device, consisting of a superfluid weak link and a mechanical element. The superfluid motion in this device is quantized. The resulting discrete energy levels are resolvable at millikelvin temperatures essential to maintaining the superfluid state. Appropriate device engineering can yield the necessary nonlinearity to realize qubit functionality. Hence, this device can potentially operate as a charge-neutral, superfluid quantum bit with  micron-sized dimensions and millisecond scale coherence time. We show that this quantum regime is within reach for a range of device designs. 
\end{abstract}

\maketitle

\section*{Introduction}
In the pursuit of hardware for scalable quantum information processing, a variety of natural and engineered systems are currently under exploration. These include architectures based on photons \cite{slussarenko2019photonic,O_Brien_2009,Flamini_2018}, ions \cite{Bruzewicz_2019,HAFFNER_2008,monroe2013scaling}, atoms \cite{henriet2020quantum,saffman2016quantum} and impurity systems \cite{Zhang_2020} as well as solid-state  quantum dots \cite{Vandersypen2019a,Chatterjee_2021,Burkard_2023} and superconducting qubits \cite{Bravyi_2022,Kjaergaard_2020}. 
In this work, borrowing concepts from superconducting qubits, we propose an architecture based on the charge-neutral superfluid $^3$He and the quantized states it exhibits in a suitably engineered device. While the key idea of using Josephson junctions transfers to the world of $^3$He \cite{RMPJosephson:2002}, charge neutrality leads to fundamental differences in the interaction with external electromagnetic fields when compared to the electronic Cooper pair condensate of a superconductor. As a result, flux and charge noise may not pose a challenge for devices based on neutral superfluids. In addition to the Josephson effect, the device relies on mechanical potential energy, which plays a similar role to the charging energy in superconducting circuits. We discuss the design of  a Superfluid Helium Oscillator Quantum (SHOQ) device and propose its operation in a regime where it may maintain phase coherence for times long enough to function as a qubit. 

The superfluid phases of $^3$He are characterised by a  $p$-wave spin triplet order parameter, dissipationless superfluid flow and a gap in the quasiparticle spectrum. In this work, we focus on the case of superfluid $^3$He in the B-phase. The superfluid B-phase is an equal mixture of all three spin states of the spin triplet manifold \cite{BW}. In the  B-phase, the superfluid gap is isotropic across the spherical Fermi surface, similar to the superconducting gap in BCS superconductors. A small constriction connecting two reservoirs of superfluid $^3$He can effect superfluid Josephson physics \cite{RMPJosephson:2002}. The required size of the constriction, 
or ``weak link", is set by the superfluid coherence length, which varies from $\sim 10-80$\,nm for the range of hydrostatic pressures in the B-phase. 
Weak links in superfluid $^3$He have been realised experimentally using nanoapertures as pinholes linking superfluid reservoirs \cite{RMPJosephson:2002}. Some of these experiments report sinusoidal relations between mass-current and phase. Josephson-like coupling between two samples of $^3$He depends critically on the symmetry and texture of the order parameter in the samples on the two sides of the link.  
Unusual non-sinusoidal Josephson relations between mass current and phase can arise due to the internal spin structure of Cooper pairs \cite{ThunebergReview:2005}. 
However in all cases, this current-phase relation, including cases with textural dissipative effects, is nonlinear. This facilitates its function as a nonlinear element in the  SHOQ device. 
\setcounter{figure}{0}
\begin{figure}
\includegraphics[width=0.4\textwidth]{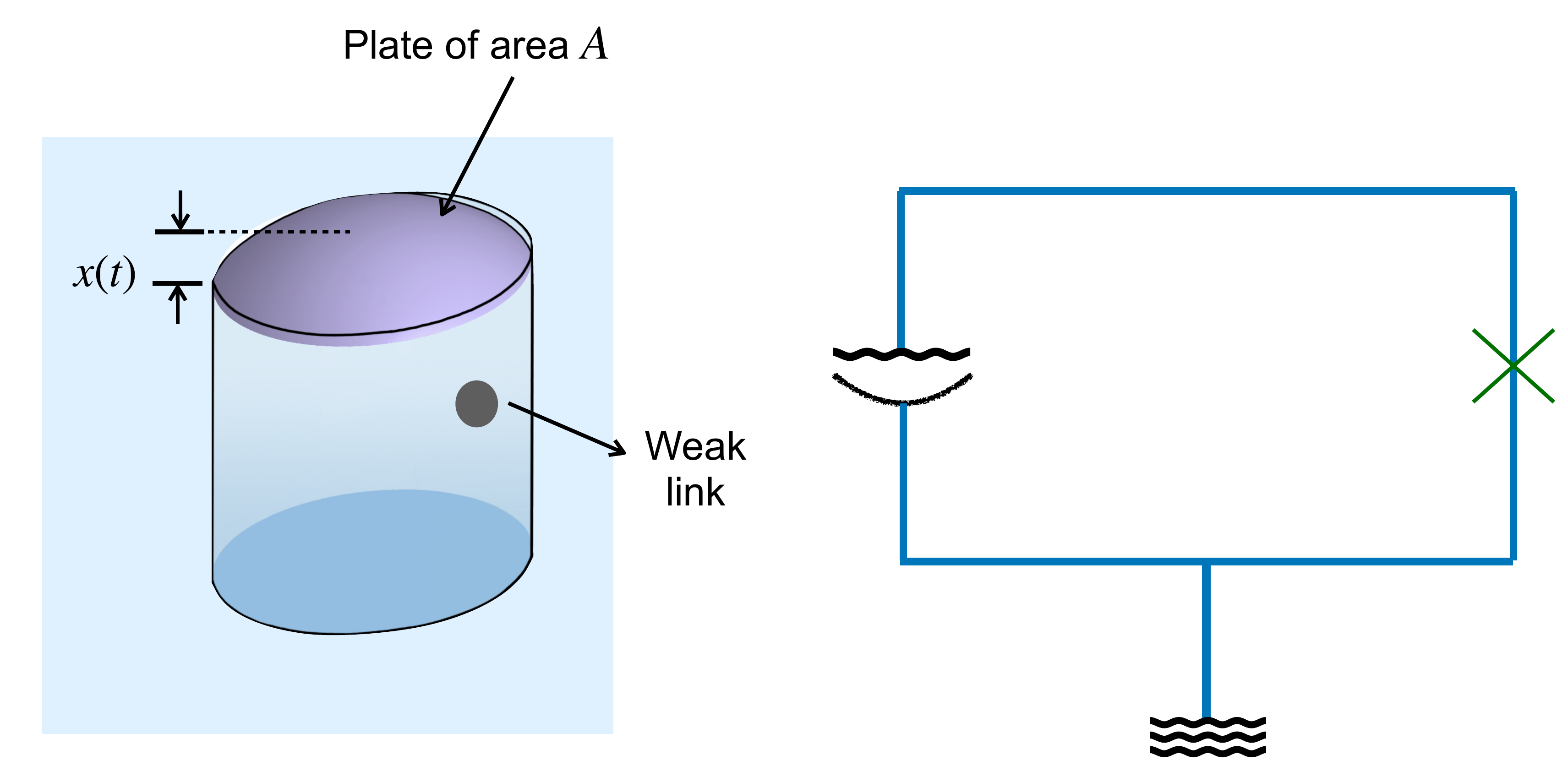}
    \caption{Left panel: Schematic of the proposed SHOQ device. The device is composed of a cylindrical cell with a small aperture that functions as a superfluid weak link. The lid of the cell is an elastic plate that displaces as the pressure inside the cell changes. The displacement of the plate $x(t)$ is not shown to scale. Right panel: Equivalent circuit diagram for the SHOQ device. The superfluid ``circuit'' contains a weak link denoted by $\times$ shunted by the elastic plate element that functions as a fluidic capacitor. The  arc in the symbol for the fluidic capacitor refers to the mechanical plate element and the wiggly line refers to the coupled superfluid.  The bottom circuit symbol denotes the superfluid ``ground''. The blue connections represent flow of superfluid mass current. }
    \label{fig:CellDesign}
\end{figure}

{A constriction connecting two reservoirs of superfluid $^4$He needs to be atomically small to realise Josephson physics. This has been demonstrated at temperatures critically close to the superfluid $^4$He transition temperature \cite{Whistling} where the superfluid coherence length gets larger. In this temperature regime, superfluidity is suppressed and the normal fluid component also contributes to the kinetic inductive energy, in addition to the superfluid contribution. This is unlike the superfluid $^3$He case and the Josephson relations for superfluid $^4$He have not been well-studied due to  experimental challenges. We focus here on superfluid $^3$He.}

Based on several prior works \cite{RMPJosephson:2002,ThunebergReview:2005}, we assume that nanoapertures function as weak links with sinusoidal current-phase relations.  We assume smooth and uniform textures of superfluid $^3$He-B on each side of the weak link and across the aperture straddling both sides. We disregard the role of surface roughness, disorder and other factors that affect the order parameter internal structure/orientation \cite{SaulsJosephson:1990,ViljasThuneberg:2004,ViljasThunebergPRL}, as well as texture-related effects including dissipative effects. We justify this assumption in Methods.
We consider the simple case of a pristine superfluid enclosed by smooth surfaces, noting that the role of textures, surface roughness and the choice of superfluid phase may lead to additional fascinating effects beyond the scope of this paper.  
\section*{Results}
\subsection*
{The Superfluid Helium Oscillator Quantum (SHOQ) Device}
\label{sec:cell}

We design a device that consists of  a cylindrical cell with a Josephson weak link. A schematic of the device is shown in Fig. \ref{fig:CellDesign}. The cell dimensions are large compared to the superfluid coherence length, $\xi$ 
The cell contains superfluid $^3$He-B and is coupled to a $^3$He-B reservoir via an aperture (acting as a Josephson weak link when of size $\sim \xi$). The cell is immersed in superfluid. The superfluid phase difference $\varphi$ across this weak link induces a Josephson current through the weak link 
. One  surface of the cell is an elastic plate that can be displaced or deformed as the pressure inside the cell changes. All other walls of the cell are taken as rigid and fixed. This can  be achieved by carving out a cavity inside a stiff bulk material such as quartz. This provides a rigid frame to which one attaches a flexible plate or membrane (a thin wafer of a chosen material). 

We consider  a superfluid phase difference $\varphi \neq 0$  established across the weak link at time $t=0$. A Josephson  current $I$ flows through the weak link creating a pressure difference $\delta P$ between the the superfluid in the cell and that in the reservoir. This pressure difference is also the pressure difference between the two sides of the plate. This $\delta P$ provides the driving force for small displacements, $x(t)$ of the plate from equilibrium. The stiffness of the plate provides a restoring force that relaxes the plate back to its equilibrium position. 
This in turn, drives a mass current through the weak link to restore the pressure difference across the weak link. The device hosts superfluid oscillations across the weak link, accompanied by  oscillations of the plate. 

We denote the energy stored in the phase difference across the weak link by $E_W$.  We denote the  elastic energy stored in the plate by $E_P$. The total energy stored in the device oscillates between  $E_P$  and  $E_W$ in each cycle. We assume 
that the fluid is incompressible (justified in Methods).

The Josephson-Anderson phase evolution equation relates the rate of change of $\varphi$ to the potential difference $\delta \mu$ across the weak link (see Methods for a brief review),
\begin{equation}
\label{PhaseEvolutionEqn}
    \frac{d\varphi}{dt} = -\frac{\delta\mu}{\hbar}\,\,\,.
\end{equation}
For a superfluid, variations of the potential $\delta \mu$ are given by
\begin{equation}
\label{muP}
    \delta\mu = \frac{2m}{\rho}\delta P\,\,,
\end{equation}
where $m$ is the mass of $^3$He quasiparticles and $\rho$ is the superfluid density.
$E_W$ is  the work done by the mass current flowing through the weak link i.e., the time integral of the mechanical power of the mass current flowing through the weak link. Analogous to  electrical power which is the product of the electrical potential difference and the electrical current, the mechanical power is given by $\frac{\delta\mu}{2m}\, I$ \cite{RMPJosephson:2002}. 
$E_W$ is then given by
\begin{equation}
\label{powerEW}
    E_W =  \int_0^t dt'\, \frac{\delta \mu}{2 m}\, I\,\,\,.
\end{equation} 
Using equation (\ref{PhaseEvolutionEqn}), this yields
\begin{eqnarray}
\label{EW}
    E_W &=& - \int_0^t dt'\,\frac{\hbar}{2 m}\,\frac{d\varphi}{dt'}\,I =   \int_0^\varphi \frac{\hbar}{2 m}\, I(\varphi')d\varphi'\\
    \nonumber
    &=& -\frac{\hbar}{2m} \,I_c\,\cos\varphi\,\,\,,
\end{eqnarray}
for a sinusoidal Josephson relation, $I(\varphi) = I_c \sin\varphi$, where $I_c$ is the critical current. 
{For micron-sized devices, the kinetic energy of superflow through the weak link $E_W$ dominates the energy associated with bulk superfluid flow within the device, $E_{bulk}$. $E_{bulk}/E_W \ll 1$ and  $E_{bulk}$ may be neglected. We assume the weak link is a pinhole (with negligible width) and there is no kinetic inductance associated with superflow through the pinhole channel. }

We consider the motion of the plate, of area $A$, to derive $E_P$. For small plate displacements, the superfluid functions as the mass and the  plate as the spring in a mass-spring system with a spring constant or stiffness constant $k$. The plate displacement is linear in the pressure difference, $\delta P = (k  x(t))/A$ and the plate responds as a simple harmonic oscillator. Similar designs have been experimentally demonstrated 
in centimeter-sized  cells for the study of superfluid weak links \cite{RMPJosephson:2002}; and more recently, in millimeter-sized Helmholtz resonator cells for the study of superfluid helium physics \cite{Schwab:2017,SourisDavis:2017}.  We write $\dot\varphi$, the rate of change of $\varphi$  described by Eq.\ \eqref{PhaseEvolutionEqn} in terms of $x(t)$  ,
\begin{equation}
\label{JosephsonEquation2}
\dot{\varphi} \equiv \frac{d\varphi}{dt} = -\frac{\delta\mu}{\hbar} = - \frac{2m\delta P}{\rho\hbar} = - \frac{2mk}{\rho\hbar \,A}x(t)\,\,\,.
\end{equation}
The energy stored in the plate, $E_P$ is the sum of kinetic and potential energies associated with the simple harmonic motion of the plate. Typically, the mass of the plate is much smaller than the mass of helium in the cell (further justified in Discussion) and the potential energy of the plate dominates. Then, the kinetic energy of the plate may be disregarded and $E_P$ is given by 
\begin{equation}
\label{EP}
E_P = \frac{1}{2} k x^2 =  \frac{\rho^2\hbar^2 A^2}{8 k m^2} \dot{\varphi}^2\,\,\,,
\end{equation}
using equation(\ref{JosephsonEquation2}). 

Examining $E_W$ and $E_P$ in equations (\ref{EW}) and (\ref{EP}) respectively, we identify  $\varphi$ as the dynamical variable in this device. 
We choose $E_W$ as the potential energy and $E_P$ as the kinetic energy in the device.  We  construct the device Lagrangian in terms of $\varphi$ and $\dot{\varphi}$, $\mathcal{L} = E_P - E_W$,
\begin{equation}
\label{Lagrangian}
    \mathcal{L} =  \frac{\rho^2\hbar^2 A^2}{8 k m^2} \dot{\varphi}^2 + \frac{\hbar}{2m} I_c \cos\varphi
    \,\,\,.
\end{equation}
 The quantum nature of the variable $\varphi$  allows us to quantize the SHOQ device using  conventional canonical quantization procedure, following the well-established formulation for superconducting quantum circuits \cite{WendinReview:2017}.


\begin{center}
\begin{table*}[!ht]
\begin{tabular}{c|c|c}
Physical quantity & Superconducting circuit & SHOQ device\\
\hline
Chemical potential difference & $V$(voltage) & $\frac{2\pi}{\rho\kappa_0} \delta P$ \\
Current & Electrical current & Mass current\\
Coordinate variable & Generalized flux ($\Phi$) & Generalized circulation ($\mathcal{K}$)\\
Momentum variable & 
$Q_e = 2en_e$ & 
$Q = 2mn$\\
Fundamental Quanta & Flux Quantum $\Phi_0 = \frac{h}{2e}$ & Circulation Quantum $\kappa_0 = \frac{h}{2m}$\\
Conjugate variables & ($\Phi$,$Q_e \propto\dot{\Phi}$) & ($\mathcal{K}$,$Q\propto\dot{\mathcal{K}}$)\\
Number operator($\hat{n}$) & $\frac{Q_e}{2e}$ & $\frac{Q}{2m}$ \\
$E_C$ & $\frac{e^2}{2 C}$  & $\frac{k m^2}{2\rho^2 A^2}$ \\
$E_J$ & $\frac{\hbar}{2e}I_c^e$  & $\frac{\hbar}{2m}I_c$\\
\hline
\end{tabular}
\caption{Glossary of physical quantities in the SHOQ device and their respective analogues in a superconducting quantum  circuit. $e$ is the charge of the electron. $C$ refers to the net  electrical capacitance in the superconducting circuit.  $I_c^e$ is the critical electrical current in the Josephson junction in the superconducting circuit. $n_e$ is the number operator in the superconducting circuit quantization scheme.}
\label{tab:Glossary}
\end{table*}
\end{center}
\subsection*
{SHOQ Circuit Theory}
\label{sec:ckt-thry}
The Hamiltonian formulation for the dynamics of electrical circuits, along with its quantum description is firmly established as the backbone of circuit theory for superconducting quantum circuits \cite{Devoret:1997,WendinReview:2017}. 
We extend this theory to the dynamics of  super\emph{fluid}  circuits such as the SHOQ circuit.  In order to make the analogy with  superconducting circuit theory, we  invoke the concept of ``circulation'' in the context of superfluids (reviewed in Methods).
We define the {\emph{generalized}} circulation, ${\mathcal{K}}$  analogous to the generalized flux in the superconducting circuit case.
 $\mathcal{K}$ 
is defined as the time integral of the instantaneous pressure difference across the weak link:
\begin{equation} 
    \mathcal{K}(t) = \int_0^t dt'\,\frac{\delta P(t')}{\rho}\,\,\,.
\end{equation}
We integrate the Josephson equation (\ref{JosephsonEquation2}) and express it  in terms of the circulation quantum $\kappa_0 \equiv \frac{h}{2m}$, 
\begin{equation}
    \label{phidotkappadot}
     \varphi(t) = -\frac{2\pi}{\kappa_0} 
    \mathcal{K}(t) \,\,\,.
\end{equation}
We can now write the 
the Lagrangian (\ref{Lagrangian}) in terms of $\mathcal{K}$,
\begin{equation}
\label{LagrangianK}
\mathcal{L} =
    \frac{h^2 \rho^2 A^2 }{8 k m^2 } \frac{\dot{\mathcal{K}}^2}{\kappa_0^2} + \frac{\hbar}{2m} I_c \cos(2\pi\frac{\mathcal{K}}{\kappa_0})\,\,\,.
\end{equation} 
The equation of motion for $\mathcal{K}$ is
\begin{equation} 
\ddot{\mathcal{K}}  =  -\frac{\kappa_0}{2\pi}\omega_p^2\,\sin(\frac{2\pi}{\kappa_0} 
    \mathcal{K})\,\,\,,
\end{equation}
where $\omega_p$ is given by
\begin{equation}
 \omega_p^2 = \frac{2mk}{\rho^2 A^2\hbar}\,I_c\,\,\,,
\end{equation}
using equations (\ref{JosephsonEquation2}) and (\ref{phidotkappadot}) (shown explicitly in Methods). 
The SHOQ circuit shown in Fig.\ \ref{fig:CellDesign} is now described by the Lagrangian in equation (\ref{LagrangianK}) with the circuit degrees of freedom
${\mathcal{K}}$ and $\dot{\mathcal{K}}$. 

The canonical momentum, $Q$ associated with the dynamical variable $\mathcal{K}$ is given by
\begin{equation}
\label{Qdefn}
    Q \equiv \frac{\partial \mathcal{L}}{\partial\dot{\mathcal{K}}} = 
    \frac{ (\hbar  \pi \rho A)^2 }{k m^2} \frac{\dot{\mathcal{K}}}{\kappa_0^2}.
\end{equation}
Now  defining $n 
\equiv Q/(2 m)$, 
the Hamiltonian is,
\begin{equation}
\label{HamiltonianSuperfluid}
    \mathcal{H} 
    = \frac{2 km^2}{\rho^2 A^2} n^2 - \frac{\hbar}{2m} I_c \cos(2\pi\mathcal{K}/\kappa_0) \,\,\,.
\end{equation}
Following canonical quantization procedure, we promote $Q$ and $\mathcal{K}$ to operators such that $[\hat{\mathcal{K}}, \hat{Q}] = i\hbar$.
{We interpret $\mathcal{K}$ as the  number of circulation quanta and $n$ as the number of Cooper pairs transferred across the weak link in the SHOQ device. }
We compare equation (\ref{HamiltonianSuperfluid}) to the superconducting transmon (or Cooper Pair Box)  Hamiltonian $\mathcal{H}_{SC}$,
\begin{equation}
    \mathcal{H}_{SC} = 4\mathcal{E}_C n^2 - \mathcal{E}_J \cos\phi\,\,\,,
\end{equation}
where $\mathcal{E}_C$ and $\mathcal{E}_J$ are the transmon charging energy and Josephson energy respectively; $\phi$ is the phase difference of the superconducting wave function across the Josephson junction in the transmon circuit.
We identify superfluid analogues $E_C$ and  $E_J$,
\begin{equation}
\label{ECJanalogues}
    E_C = \frac{k m^2}{2 \rho^2 A^2}\,\,\,;\,\,\,E_J = \frac{\hbar}{2m}I_c\,\,\,,
\end{equation}
where $E_C$ is the energy associated with the equivalent fluidic capacitance \cite{RMPJosephson:2002} of the plate and $E_J$ is the superfluid Josephson energy.
The oscillator frequency $\omega_p$  is given by $\hbar\omega_p = \sqrt{8E_J E_C}$ in equation (\ref{ECJanalogues}) above, just as for the superconducting circuit case. We point out that for small $\varphi$, equation (\ref{HamiltonianSuperfluid}) is the Hamiltonian for a quantum harmonic oscillator.

We equip ourselves with a dictionary of  quantities in the SHOQ device that correspond to familiar analogues in the superconducting quantum circuit.
 We summarise these analogues in Table \ref{tab:Glossary}. 

\subsection*
{Towards a Quantum Regime for a Qubit}
\label{sec:QuantumRegime}

We now explore if the SHOQ circuit can be designed in a reasonable operating range to realize the basic criterion of having energetically distinct qubit states sufficiently apart so that the lower state may be preferentially occupied at a chosen operating temperature. 
At a pressure of $P = 21$ bar, the superfluid transition temperature is $T_c = 2.27$ mK. The superfluid gap at low temperatures, ${\bf{\Delta}}(T,P)/h < {\bf{\Delta}}(T=0,P)/h = 87.47\,$MHz 
at $P = 21$ bar \cite{Spindry}.
We design a SHOQ device with a characteristic frequency, $\hbar\omega_p < {\bf{\Delta}}$. We use the known values of the effective mass and density for liquid $^3$He \cite{RMPWheatley:1975} at 
$P = 21$ bar  
Theoretical estimates for the  critical areal current density are $I_c \sim 1$ kg \,\,m$^{-2}$ s$^{-1}$ \cite{RMPJosephson:2002}. 
We consider a typical weak link aperture of size $10$ nm $\times 10$ nm  similar to apertures used in superfuid weak link experiments  \cite{RMPJosephson:2002}. 
For the plate element in the SHOQ device, we consider a circular disk of radius
$5 \mu$m.
We use a nominal value of $k = 10^7$ N/m which is a good estimate for materials used in superfluid weak link experiments \cite{DavisEmail,RMPJosephson:2002,SourisDavis:2017}. 
We recognize that this design achieves
   $ T < \omega_p/2\pi < \Delta$
if the SHOQ device is operated at a temperature $T \lesssim 0.8$ mK. 
In this regime, the eigenstates of the SHOQ circuit Hamiltonian are quantized with the ground and excited states separated by a frequency of 
$\omega_p/2\pi = 14.25$ MHz. These levels are distinct and and 
the thermal occupation of the first excited state is 
$10.3$\% at a temperature of
$T = 0.3$ mK. The superfluid state is robust to excitations of energy $\omega_p/2\pi < {\bf{\Delta}}$ and we thus show that the quantum regime is indeed attainable with a SHOQ circuit.

For the design suggested above to potentially function as a qubit, we require that the spacing of SHOQ energy levels be non-uniform to enable control and manipulation of the qubit by reducing the risk of unwanted transitions to higher SHOQ levels. We use the definitions of   anharmonicity from the transmon analogy \cite{Koch:2007} and examine the anharmonicity of the SHOQ device. The absolute anharmonicity $\alpha$ is  given by $E_C/h = 99.62$ kHz. The relative anharmonicity $\alpha_r = \alpha / \hbar\omega_p = \sqrt{E_C/(8E_J)}$.
For the design above, $E_J/E_C = 2.56 \times 10^3$ and $\alpha_r = 0.007$.
We provide a range of example SHOQ device designs and list their operating parameters in Table \ref{tab:SHOQDesigns}.

In some designs the anharmonicity of the SHOQ device is rather week. In the electrical analogy, these designs are circuits that operate very deep in the transmon regime. In comparison with the transmon, the weakness of the anharmonicity in the SHOQ circuit is related to the choice of device parameters, constrained to realistic values for the designs in Table \ref{tab:SHOQDesigns}.

 Some designs in Table \ref{tab:SHOQDesigns}  could realise sufficient anharmonicity in comparison to transmons, confining probability amplitudes of state occupation to the lowest two qubit states. Reducing the area of the plate $A$ as well as using stiffer materials with larger $k$, increases $\omega_p$ and increases $E_C$, while decreasing $E_J/E_C$ just as we need to realise a SHOQ bit. However, this direction needs to be pursued with caution as progressively smaller plates will lead to effects arising from superfluid confinement. 
 
\begin{figure}
\includegraphics[width=0.3\textwidth]{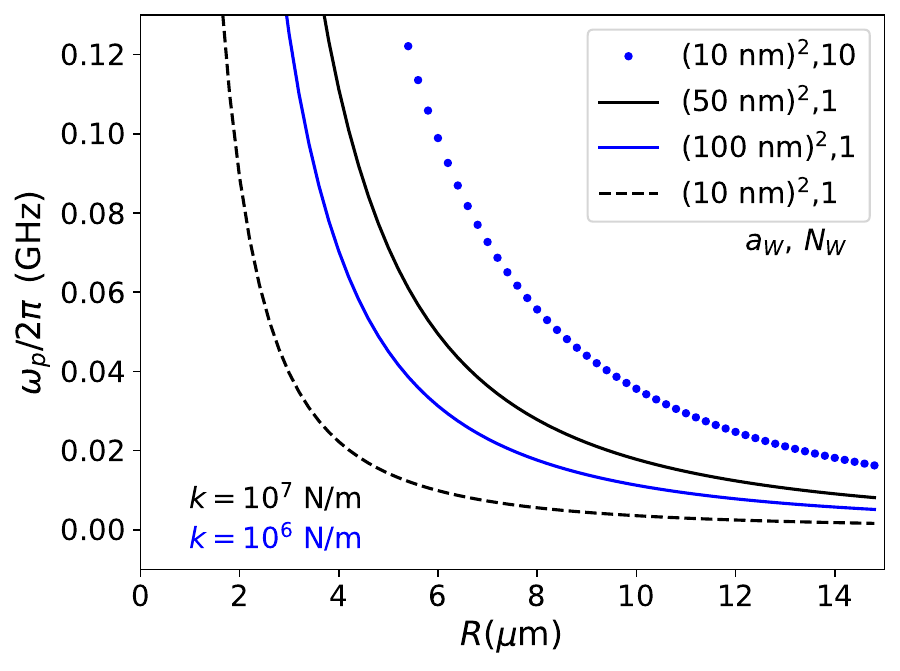}
\includegraphics[width=0.3\textwidth]{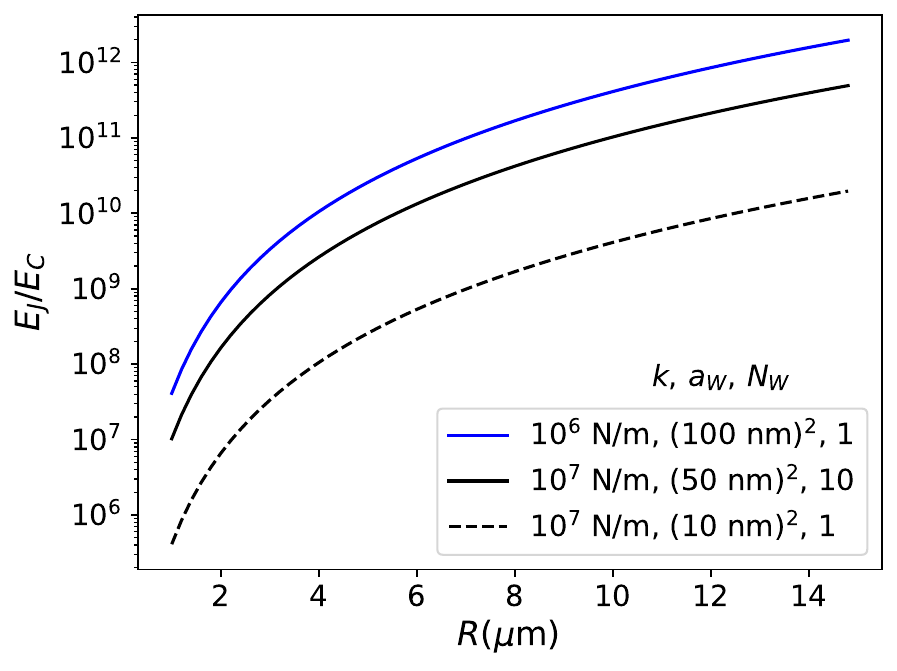}
      \caption{Dependence of operating parameters $\omega_p$, $E_J/E_C$ on design specifications - $R$, aperture dimensions, $k$ at pressure $P = 21$ bar. $a_{W}$ and $N_{W}$ are, respectively, the area of the nanoaperture and the number of nanoapertures that functions as the weak link in the design. $R$ is the radius of the plate element in the SHOQ device..}
      \label{fig:plots}
\end{figure}

 In Figure \ref{fig:plots}, we visualize the dependence of various operating parameters on design specifications.

\begin{center}
\begin{table*}
\begin{tabular}{c|c|c|c|c|c|c|c|c|c}
\hline
Index & $k$(N/m) & $a_{W}$(nm$^2$) & $N_{W}$ & $R (\mu$m$)$  & $\omega_p/2\pi$ (MHz)  & $E_C$ (kHz) & $\frac{E_J}{E_C}$ & ${T}_{sqt}@0.4$ mK (ms) & ${T}_{sqt}@0.3$ mK (ms)\\
\hline
\#1&$10^6$ & $100\times 100$ & $1$ & $8$ & $17.60$ & $1.52$ & $1.68 \times 10^7$ & $0.57$ & $23.35$\\
\#2&$10^7$ & $100 \times 100$ & $1$ & $5$ &  $142.48$ & $99.62$ & $2.56 \times 10^5$ & $0.20$ & $8.21$\\
\#3&$10^6$ & $10\times 10$ & $10$ & $5$ & $14.25$ & $9.96$ & $2.56 \times 10^5$ & $0.64$ & $25.96$\\
\#4&$10^7$ & $10 \times 10$ & $100$ & $8$  & $55.66$ & $15.20$  & $1.68\times 10^6$ & $0.32$ & $13.13$\\
\#5&$10^7$ & $10\times 10$ & $1$ & $5$ & $14.25$ & $99.62$  & $2.56\times 10^3$ & $0.64$ & $25.96$\\
\#6&$10^6$ & $10\times 10$ & $1$ & $2$ & $28.16$ & $389.06$ & $6.55 \times 10^2$ & $0.45$ & $18.47$\\
\hline
\end{tabular}
\caption{Examples of SHOQ circuits designed to operate in the quantum regime. $a_{W}$ and $N_{W}$ are, respectively, the area of the nanoaperture and the number of nanoapertures that functions as the weak link in the design. $R$ is the radius of the plate element in the SHOQ device. ${T}_{sqt}$ is the  coherence time estimated from the single quasiparticle tunneling dissipation rate discussed in the text.}
\label{tab:SHOQDesigns}
\end{table*}
\end{center}

\subsection*
{Coupling to a Superconducting Qubit}

In order for qubit function to be realised as such,  interaction of the SHOQ two-level system with external fields and/or other hybrid circuit elements is essential. Various forms of interaction  and coupling schemes provide the means for readout, state preparation, control and qubit operations for the SHOQ circuit. We suggest a quantum electromechanical scheme using hybrid circuit elements as an illustrative coupling scheme for the SHOQ bit. Hybrid systems involving disparate degrees of freedom have been employed successfully to materialise mechanical quantum ground states \cite{ClelandSET:2010,Teufel:2008}, cavity optomechanics with diverse applications \cite{Schwab:2003,SourisDavis:2017,Harris:2008}, electrons on the surface of helium for quantum computing \cite{Kawakami:2024,Pollanen:2024} among many others. 

\begin{figure}
\includegraphics[width=0.4\textwidth]{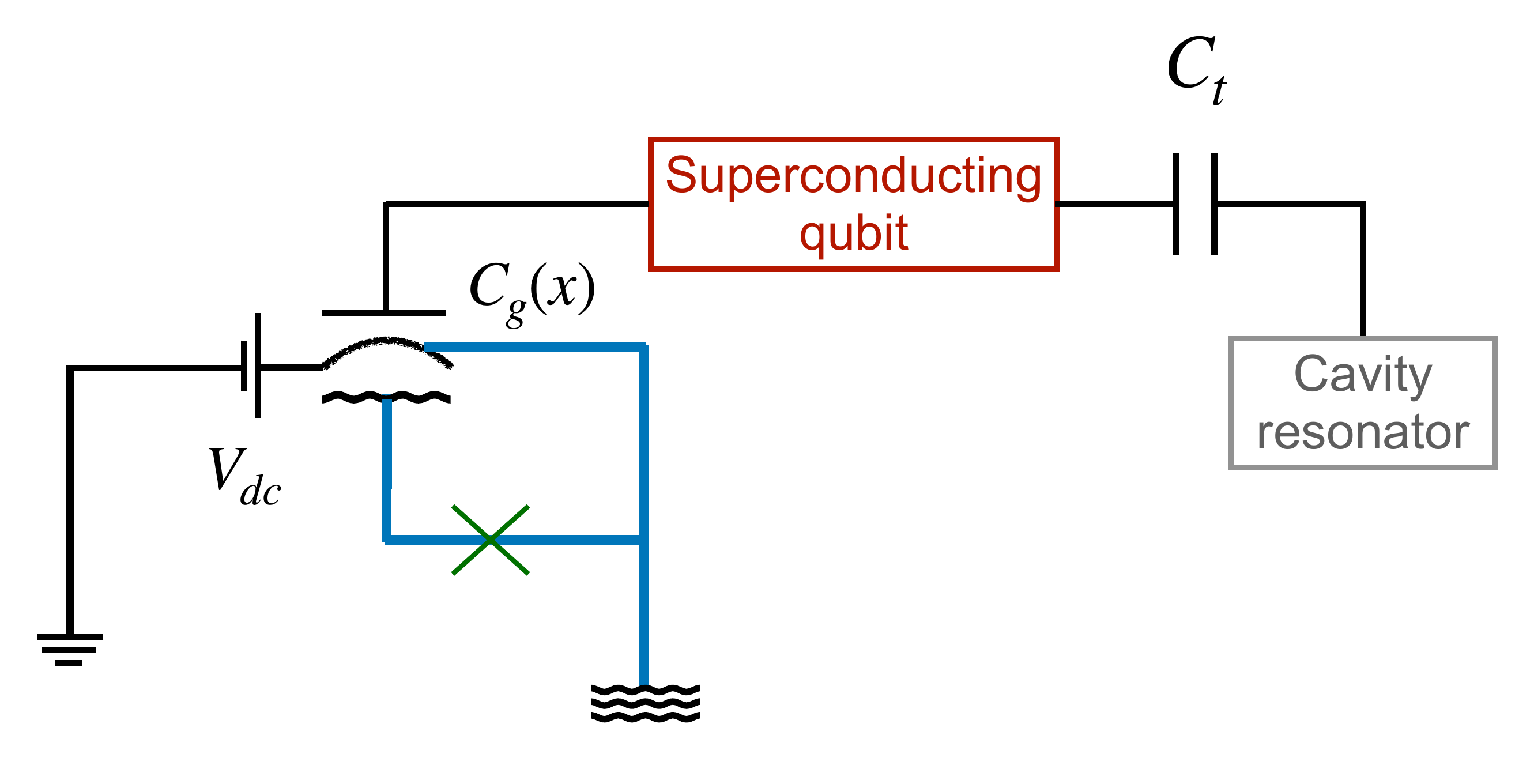}
\caption{Hybrid cavity-transmon-SHOQ bit coupling scheme. The cavity and qubit are connected via a capacitor $C_t$ as in \cite{HybridCouplingNature:2013}. 
The qubit is connected to the SHOQ device through a gate capacitance, $C_g(x)$ described in the text.
The electromechanical coupling is turned on via a gate voltage, $V_{dc}$.}
\label{fig:HybridCoupling}
\end{figure}

In hybrid circuit cavity quantum electrodynamics with mechanical elements \cite{HybridCouplingNature:2013}, coupling of phonon modes in a micromechanical resonator to both a microwave cavity and a superconducting (SC) transmon qubit (SC qubit) has been demonstrated to enable phonon-photon state transfer and to harness the mechanical degree of freedom as a qubit \cite{MechanicalQubit2024}. We suggest using a similar hybrid scheme to achieve coupling between the SHOQ device to conventional (electrical) circuit degrees of freedom. We consider a circuit composed of the SHOQ device, a transmon qubit and a microwave cavity as shown in Fig. \ref{fig:HybridCoupling}. The SC qubit would couple to the SHOQ device via a gate capacitance $C_g(x)$. With a constant gate voltage, $V_{dc}$ applied across the elastic plate element of the SHOQ device as shown in Fig. \ref{fig:HybridCoupling}, the SHOQ oscillations give rise to a ``motional gate charge'' \cite{HybridCouplingNature:2013} on the plate (see Methods)
This is an electrical charge induced on the capacitor in Figure \ref{fig:HybridCoupling} by the  motion of the 
{elastic plate component of the SHOQ device.} 
The motional gate charge gives rise to a coupling term between the SC qubit and the SHOQ device, as shown in Methods. 
{Based on the electrical parameters used in an analogous mechanical coupling scheme reported in \cite{HybridCouplingNature:2013}, we find the magnitude of this coupling to be $\sim 0.18$ MHz.  This coupling is strong enough to measure the SHOQ bit state via dispersive readout of the SC qubit state.}
The SC qubit may be used for quantum tomography of the SHOQ bit for designs with weak nonlinearity 
and SHOQ devices with high quality factors \cite{Tomography:2013}.
SHOQ bit state measurement and state tomography are possible when the state-dependent dispersive shift of the SC qubit $g_m^2/\delta$ is made larger than the SC qubit dissipation rate for state-of-the-art transmons ($T_2>1 \textrm{ms}$). This regime may be obtained for the coupling $g_m$ when the detuning $\delta$ between the qubit and resonator is around $1 \textrm{GHz}$ or less. 

The eigenstates of the coupled SC qubit-SHOQ bit system are dressed states, i.e., states in which excitations of the SC qubit and SHOQ bit hybridize. Viewing the SHOQ bit as a weakly anharmonic oscillator, we can draw on an analogy between the coupled SHOQ bit - SC qubit system with SC qubits interacting via high-$Q$ microwave resonators. Using this analogy, the SHOQ bit-SC qubit coupling scheme can provide a means to  measure the state of the SHOQ bit. 
The architecture and operation of gates for the SHOQ bit is beyond the scope of this paper and will be discussed in future works. 

We discuss sources of dissipation based on prior works on superfluid devices and estimate dissipation rates in Methods. Based on known sources, we estimate single quasiparticle tunneling to be the most limiting source. 
{We recognise that $E_J/E_C$ for the SHOQ device seems 
larger than typical values for the transmon case.} However, $\chi_{SHOQ}$ of a few MHz  realised using the scheme above could correspond to a regime where the SHOQ device nonlinearity is much larger than its decoherence rate. This rate is set by our estimate for the decoherence time due to single quasiparticle tunneling, ${T}_{sqt}$ of a suitably designed SHOQ device, i.e., $\chi_{SHOQ}\,{T}_{sqt} \gg 1$ such as as $\#6$ of Table 
 \ref{tab:SHOQDesigns}. 
 {
 {In order to ensure that the SHOQ bit maintains coherence, we would need the 
 duration of the driving pulse to be such that $\Delta t_R \ll T_{sqt}$. For design \#6 in Table II, this amounts to $ \Delta t_R \ll 18.47$ ms. Separately, $E_C$ sets the scale for the anharmonicity of the SHOQ bit levels in analogy to the transmon.
 In order to enable the spectral resolution of the SHOQ bit levels, we would also need $\Delta t_R > 1/E_C$.  For design \#6 in Table II, this amounts to $ \Delta t_R > 2.5 \mu$s.  For resonant driving, we would thus need a driving pulse with width $18.47$ ms $> \Delta t_R > 2.5 \mu$s. For resonant driving, this would mean a Rabi frequency, $f_R$ to be bounded by $(1/T_{sqt} \sim 54$ Hz$) < f_R < (E_C \sim 389$ kHz$)$. This allows a range of $t_R$ and $\Delta f_R$ that would allow Rabi oscillations and consequently, SHOQ bit behaviour in the SHOQ device. For example, $f_R \sim 1$ kHz with $\Delta t_R \sim 50 \mu$s could drive the SHOQ device with design \#6 in Table II. In order to estimate leakage from the first excited level via off-resonant driving, the probability of occupation of higher SHOQ bit states can be estimated from the analogous expression for the transmon. For design \#6 in Table II, this gives $P_{1\rightarrow 2} \sim \frac{f_R^2}{f_R^2 + \delta^2} \sim 10^{-6} \ll 1$, for detuning $\delta \sim E_C$. Therefore, we expect the SHOQ device to support near perfect Rabi oscillations in a SHOQ bit scheme.}}
The microwave cavity can be used to drive the SHOQ bit using the SC qubit as a filter and transducer between disparate degrees of freedom, through schemes similar to that illustrated in Fig. \ref{fig:HybridCoupling} using the SC qubit. Further superconducting analogues such as the superfluid analogue to the application of gate voltages to control offset charge in transmons, discussed in Methods, can prove as useful tools to drive/control the SHOQ state. 
 

The $\#5$ design in Table \ref{tab:SHOQDesigns} offers most promise as a SHOQ bit as the estimated coherence time (based on known sources of dissipation) competes with state-of-the-art fluxonium \cite{PRLMillisecQubit:2023} and transmon qubits \cite{NatureMillisecTransmon:2021} having the longest sub-millisecond/millisecond coherence times thus far. Designs  $\#1,\, \#3$ follow design $\#5$ with long coherence times. 
Making a one-to-one analogy with the transmon case, the anharmonicity is an indicator of the time taken to operate gates between transmons. Designs $\#5$ and $\#6$  offer meaningful anharmonicity to realise SHOQ bit functionality. For these designs, expected gate times (based on the analogy with the transmon) which are much shorter than both the time taken to prepare an initial state and the coherence time could be realised with the quality of the gate operation not limited by the anharmonicity.   

\section*{Discussion}
\label{sec:discussion}

The main parameters that may be engineered for the SHOQ device to operate in the quantum regime are the spring constant $k$ set by the material used for the plate element, the size of the weak link apertures that sets $I_c$ and the area of the plate $A$. 

{\it{Pressure:}} The hydrostatic pressure of the superfluid does affect material properties such as $m$ and $\rho$. However, the ratio $m/\rho^2$ that appears in $\omega_p^2$ varies by less than 4\% \cite{RMPWheatley:1975} across the entire pressure range and does not provide much utility in designing $\omega_p$. 

{\it{Plate specifications: }} The size and thickness of the plate affect the stiffness of the plate, in addition to the material it is composed of. In the MEMS geometries we suggest in our designs, these considerations become significant. We will not go into a detailed analysis of spring constants for plates made of various materials and of varying sizes. Instead, we illustrate that a workable design is achievable with known materials. The stiffest materials used as elastic membranes in similar devices are quartz and borosilicate glass with spring constants of $ 10^7$ N/m  for mm-sized disks \cite{SourisDavis:2017}. 

We have ignored the mass of the plate in our calculation and designs. The harmonic oscillations of the plate are associated with the spring constant $k$ and the associated mass is the mass of the helium in the cell. The ratio of the mass of the plate, $M_P$ to the mass of helium in the cell $M$ is $\propto \frac{\rho_{P}}{\rho} \frac{L_P}{L}$, where $\rho_P$ is the density of the plate material and $L_P$ and $L$ are the thickness of the plate and the height of the cell, respectively. For a typical quartz plate used in similar cells in  experiments \cite{SourisDavis:2017}, $L_P = 50$ nm  and the density of quartz is $\rho_P = 2.5$ g/cm$^3$. For such a plate in a cell of  minimum height $L = 10\, \mu$m, the ratio $\frac{M_P}{M} \ll 1 $ is small, with this ratio getting smaller for taller cells. Typically, the mass of the plate is much smaller than the mass of helium in the cell and ignoring the mass of the plate is a good approximation. 
{ Further, the ratio of $\omega_p$ to the natural frequency of the elastic plate is $\omega_p/\sqrt{k/M_P} \ll 1$; therefore, ignoring the harmonic motion of massive plate is a good approximation.}

{\it{Critical Current:}} The critical current $I_c$ is proportional to the cross-sectional area of the weak link aperture.
Arrays  of multiple weak link apertures have  been used in prior works, with an objective of increasing critical currents for mass current measuremenst. In a SHOQ device, such weak link aperture arrays may be used instead of a single weak link aperture to increase $I_c$ providing an additional avenue to engineer $I_c$. 
An array of $N_{W}$ identical weak links has a critical current $I_{c} = N_{W} I_c^1$ where $I_c^1$ is the critical current for a single weak link. 

{\it{Device dimensions:}} The fabrication of $\mu$m-sized devices for applications in $^3$He experiments is at the frontier of cryogenic superfluid helium technology. Experiments using such mesoscopic devices have been successful in studying novel physics of superfluid $^3$He in confinement, topological effects and have been proposed for use in dark matter detection \cite{QUEST-DMC:2024,SaundersAB:2024,ShookPRL:2024}. The confinement of superfluid on the scale of $\xi$ leads to fascinating novel physics, absent in the bulk superfluid. These include new superfluid phases \cite{VorontsovSauls:2007,Levitin:2019,DavisPDW:2020}, quantum phase transitions \cite{VorontsovSauls:2007}, half-quantum vortices \cite{AuttiHQV:2016} and inhomogeneous phases  \cite{VorontsovSauls:2007, Levitin:2019,DavisPDW:2020} among several others. These are, however, not included in our SHOQ circuit theory, which considers  superfluid in its bulk form. In order to preserve the bulk superfluid state, we refrain from choosing SHOQ device dimensions comparable to $\xi$.
We have, therefore, assumed an elastic plate element of radius of a few $\mu$ms $\gg \xi$
in our estimate above. 

We also point out that the height of the cylinder in our cell design does not appear in the SHOQ circuit Hamiltonian. This is a free parameter as such in the design thus far; keeping in mind that we require this dimension to be in the range of $2-20 \mu$ms so as to be both large enough to steer clear of confinement effects and smaller than textural healing lengths to avoid textural dissipative effects. 


{\it{Temperature:}} It is important to point out that while superfluid properties vary only weakly with pressure, the superfluid transition temperature $T_c$ and consequently the superfluid gap ${\bf{\Delta}}$ increase with pressure. This provides an avenue to engineer higher $\omega_p$ for higher pressures.
Pushing the cryogenic frontier to lower  temperatures will enable realistic operating regimes at lower pressures. Temperatures in the range of $0.4$\,mK have been reported several times in superfluid $^3$He experiments. 

{\it{Sources of Dissipation:}} Superfluid resonators with elastic elements and similar schematics, but of much larger dimensions ($mm$-sized) than those proposed for the SHOQ device ($\mu m$-sized) have been used in   experiments investigating superfluid phenomena. These resonators operate in the classical regime in these   experiments, which have mapped out the phase diagram under confinement \cite{PDWinHe3_2020} and explored aspects of optomechanics  and acoustomechanics \cite{SuperfluidOptomechanics_2021} involving coupling of superfluid collective modes to cavity modes. There also exist proposals to use the resonator as a gravitational wave detector \cite{SuperfuidGWDetector_2021}, among others. Our device resembles superfluid Helmholtz resonators \cite{SourisDavis:2017,Schwab:2017}
for which  dissipation effects have been analysed  and  $Q$-factors determined 
via a rigorous dissipation model \cite{SourisDavis:2017}. We apply estimates from these works for superfluid $^3$He in the SHOQ device in Methods. We consider the weak-link dissipation based on prior studies on superfluid $^3$He weak-links \cite{RMPJosephson:2002}. 

From our estimates for the dissipation rates arising from viscous damping, the mechanocaloric effect, damping due to two-level systems in the substrate, single quasiparticle tunneling and  weak-link dissipation, we find that the dominant dissipation channel at low temperatures is  from single quasiparticle tunneling (sqt) through the superfluid weak link. 
At the operating temperature of $0.4$ mK, we find $Q_{sqt} \sim  10^{6}$ for the \#1 design in Table \ref{tab:SHOQDesigns}.
Based on this $Q \sim Q_{sqt}$, the relaxation time $T_1 \sim T_{sqt}$  of a putative SHOQ bit
, ${T}_{sqt} = Q_{sqt}/\omega_p$
for various designs in Table \ref{tab:SHOQDesigns}, if sqt is the dominant dissipation channel. In analogy with the superconducting case, if $E_C$ sets the scale for gate operations, all designs in Table {\ref{tab:SHOQDesigns}} access a range of qubit operation where $E_C \, 2\pi\,{T}_{sqt} > 1$. 
This means that with estimates for sqt being the dominant dissipative channel, the SHOQ device could maintain coherence for long enough times to be accessible for operation as a putative SHOQ bit.  

On the quantum information processing front, recent resurgent interest in experimental realisations of superfluid $^4$He weak links suggests an optimistic outlook towards superfluid quantum circuits at more experimentally feasible  temperatures. Liquid $^4$He undergoes a phase transition to the superfluid state at a temperature of $2.17$\,K, three orders of magnitude larger than the superfluid transition temperature for superfluid $^3$He. The characteristic superfluid  coherence length in the low-temperature limit is set by the atomic spacing. However, the coherence length gets progressively larger and diverges as the temperature approaches the superfluid transition temperature. The Josephson effect in superfluid $^4$He has been observed with nanoaperture weak links at higher temperatures close to the critical temperature for superfluidity \cite{Hoskinson:2005}. We refer the reader to \cite{Sato2012,Sato2019} for a review of Josephson effects in superfluid $^4$He. 
 If weak links with superfluid $^4$He are realised in the low-temperature limit, the SHOQ circuit formulation developed in this paper is applicable with the appropriate superfluid parameters for $^4$He.
 The aperture size to realise weak links is two orders of magnitude smaller in $^4$He than in $^3$He, making weak links in superfluid $^3$He far easier to fabricate than in $^4$He. New materials  that could potentially realise Josephson physics in $^4$He could open the door to SHOQ bits at higher operating temperatures than the scheme proposed in this work.

The motivation to design a SHOQ bit is manyfold. Primarily, it is the first quantum device proposal in a charge neutral condensed matter environment. 
Further, the exotic properties of the superfluid phases of $^3$He can influence and potentially enhance the physics that underpins the SHOQ device. The superfluid phases of $^3$He are a paradigm for spontanenous symmetry breaking and a model for unconventional pairing with $p$-wave spin triplet symmetry \cite{VW}. They include topologically nontrivial phases that host exotic physics such as Majorana states, half-quantum vortices, anomalous quantum Hall effects and offer a test-bed for cosmology in the lab \cite{VW,VolovikBook} due to their rich order parameter structure. The quantization of a composite superfluid degree of freedom as in the SHOQ device could provide unanticipated routes to exploring this exotic physics. The  Josephson effect in superfluid $^3$He provides avenues to investigate weak links between superfluids with disparate textures and further, weak links between disparate superfluid phases. The SHOQ device could potentially be explored as a probe of  such superfluid $^3$He Josephson phenomena.  The quantum superfluid circuit opens up avenues to explore both exotic physics as well as exotic quantum engineering schemes in unprecedented ways. The pristine state of superfluid $^3$He with no complications  arising from impurity
or interfacial scattering that dominate in superconducting platforms, provides unrivaled added benefits. 

\section*{Methods}
\subsection*{Homogenous Texture Across the Weak Link}

In the development of SHOQ circuit formulation in this paper, we assume smooth and uniform
textures of superfluid $^3$He-B on each side of the weak link. This assumption is correlated with the sinusoidal form of the Josephson relation we use in this work. We justify this assumption here.

The $p$-wave spin triplet nature of the superfluid $^3$He order parameter awards internal structure to Cooper pairs.
The structure of the superfluid condensate is  affected by orientational forces such as the interaction of Cooper pairs with hard boundaries or surfaces within which the superfluid is enclosed. The order parameter structure is referred to as a texture. Smooth variations of this order parameter structure evolve on length scales much longer than the superfluid coherence length. 
For weak links between superfluids in the same superfluid phase 
, Josephson relations differ for superfluids with different textures on the two sides of the link \cite{RMPJosephson:2002}. The bending energy of the order parameter is associated with changes to the internal structure of Cooper pairs in a given superfluid phase.
It is known \cite{VW} that this bending energy is minimised when textures assume the smoothest possible configuration. 
The textural healing length is the length scale over which textures return to their undisturbed (no orienting forces) state. 
The surface healing length, $\xi_S$, of textures in $^3$He-B  is of the order of  $\xi_S \sim 100 \mu\text{m}$ at temperatures of $\sim 0.5 T_c$, increasing at lower temperatures.
In containers much smaller than $\xi_S$, it is energetically more favourable to maintain a uniform texture than to adjust the internal configuration of Cooper pairs close to the surface. 
Therefore, for a cell with dimensions $\ll \xi_S$, the order parameter is oblivious to the surface and 
we may ignore the bending energy of the order parameter. This justifies our assumption of a homogeneous texture 
for devices with size dimensions much smaller than the textural healing length.
This assumption is consistent with the Josephson current being sinusoidal in the superfluid phase difference between the two sides of the weak link. 
Such a uniform texture with sinusoidal current-phase relation is known to be realised for a weak link connecting two $^3$He-B reservoirs, with an order parameter configuration of parallel orbital vectors for Cooper pairs on both sides of the weak link \cite{ThunebergReview:2005}. 


\subsection*{Josephson dynamics in superfluid $^3$He}
We briefly review Josephson dynamics in superfluid $^3$He from the perspective of applying it to the SHOQ device. In the process, we develop a ``dictionary'' (see Table  \ref{tab:Glossary}) of analogous superfluid quantities  that correspond to familiar analogues in the superconducting case.  The Josephson-Anderson phase-evolution equation is given by
\cite{RMPJosephson:2002,Hoskinson:2005},
\begin{equation}
\label{PhaseEvolutionEqn2}
    \frac{d\varphi}{dt} = -\frac{\delta\mu}{\hbar}\,\,\,,
\end{equation}
where $\delta\mu$  is the chemical potential difference and $\varphi$ the superfluid phase difference between the two sides of the weak link. We use the notation $\delta$ for variations of quantities across the two sides of the link, while using $d/dt$ differential notation for continuous changes of variables. For a fluid, chemical potential variations $\delta
\mu$ are given by
\begin{equation}
\label{dmudP}
\delta \mu = \frac{m\,\delta P}{\rho} + S \, \delta T\,\,\,,
\end{equation}
with  $S$ being the entropy and the definitions of $m$, $\rho$, $\delta P$, $T$ as used in the main text.
The temperature variations in $^3$He weak-link experiments 
are known to be negligible \cite{RMPJosephson:2002} and therefore,
\begin{equation}
\label{2dmudP}
\delta\mu = \frac{2m}{\rho} \delta P\,\,\,,
\end{equation}
since the flow is of Cooper pairs of mass $2m$.
The hydrostatic pressure difference across the plate, $\delta P$ displaces the plate by $x(t)$ and
\begin{equation}
\label{Press-x}
    \delta P = \frac{k}{A}\,x(t)\,\,\,.
\end{equation}
From equations (\ref{PhaseEvolutionEqn2})-(\ref{Press-x}), we get an equation for $\varphi$ in terms of $x(t)$,
\begin{equation}
\label{phidot}
    \dot{\varphi} = - \frac{\delta\mu}{\hbar} = - \frac{2m\delta P}{\rho\hbar} = - \frac{2mk}{\rho\hbar \,A}x(t)\,\,\,.
\end{equation}
The fluid entering the cell volume  displaces the plate such that the mass current $I$,
\begin{equation}
\label{Ixdot}
    I = \rho\,A\,\dot{x}\,\,\,.
\end{equation}
We assume a sinusoidal Josephson relation between the mass current and superfluid phase difference,
\begin{equation}
\label{sinJosephson}
    I = I_c \,\sin\varphi\,\,\,.
\end{equation}
From equations (\ref{phidot}-\ref{sinJosephson}), we obtain the equation of motion for $\varphi$,
\begin{equation}
\label{phidotdot}
\ddot{\varphi} 
= -\frac{2mk}{\rho^2 A^2\hbar}\,I_c\,\sin\varphi
\equiv - \omega_p^2\,\sin\varphi\,\,\,.
\end{equation}

\subsection*{Free Energy and Compressibility of the Superfluid in the Cell }

In Results, we assume that the fluid is incompressible and that the total free energy of the fluid in the cell is
unchanged by the plate motion as well as the mass current through the weak link. We justify this here.

As discussed in Results, the SHOQ device is designed to operate in a regime $\hbar \omega_p < {\bf{\Delta}}$. In this regime,
the plate motion (on the time-scale set by $\omega_p^{-1})$ is slow compared to microscopic time-scales relevant to the formation and stability of the  superfluid phases in $^3$He. These are given by the superfluid gap ${\bf{\Delta}}$ and the Fermi energy $\varepsilon_F$, with corresponding characteristic times $h/{\bf{\Delta}} \gg h/\varepsilon_F$.
The energy stored in the fluid in the cell, $E_{He}$, is given by thermodynamics in this regime \cite{VW}. Changes to this energy $\delta E_{He}$ contribute to the dynamics of the SHOQ system. $\delta E_{He}$ is given by 
\begin{equation}
    \delta E_{He} = - P\delta V + \frac{\mu}{2m}\delta M + T \delta S  - V\delta P + \frac{M}{2m} \delta\mu\,\,\,.
\end{equation}
Here, the flow of Cooper pairs of mass $2m$ gives rise to a change in volume $\delta V$ at pressure $P$. Changes in pressure $\delta P$ due to the motion of the plate at volume $V$ give rise to changes in the chemical potential $\delta\mu$, measured with respect to the chemical potential in the superfluid reservoir outside the cell. $M$ is the total mass of fluid in the cell. Experimental data \cite{RMPJosephson:2002} show that  temperature variations within the cell are known to be negiligible. 
Since the flow of Cooper pairs carries no entropy, we assume that $\delta S = 0$.  For the displacement $x$ of the plate, of area $A$, at a   pressure $P$ of the superfluid with density $\rho$, the change in the energy stored in the fluid is
\begin{equation}
\label{ECdot}
    {\delta E_{He}} = - PA\, {x} + \frac{\mu}{2m} \rho A \, x - V \, P + \frac{M}{\rho} \, P = 0\,\,\,,
\end{equation}
where $x$ refers to the displacement of the plate from equilibium (as in the main text), using equation (\ref{2dmudP}).

The compressibility of  superfluid $^3$He, $\Xi_h$ contributes to the motion of the plate element in the SHOQ device in Figure \ref{fig:CellDesign}. The force experienced by the plate, $F_{P}$ is the sum of the force due to the pressure difference $\delta P$ between the two sides of the plate, $F_{\delta P} = A \delta P$ and the force exerted by the compressing superfluid on both sides of the plate, $F_{\Xi} = \frac{A}{L} (\Xi_{h,in}^{-1} - \Xi_{h,out}^{-1})\, x$; $F_P = F_{\delta P} + F_{\Xi}$. Here, $\Xi_{h,in/out}$ are the respective compressibilites of the superfluid inside/outside the cell. The compressibility is given by $\Xi_h = (\rho c_1^2)^{-1}$, where $c_1$ is the speed of longitudinal (first) sound in superfluid $^3$He. For typical pressure differences $\delta P$ in the range $\lesssim 10$\,Pa as demonstrated in weak link experiments \cite{RMPJosephson:2002}, $F_{\Xi} \ll F_{\delta P}$ by several orders of magnitude and we ignore the compressibility of the superfluid in our SHOQ circuit formulation. 

\subsection*{Weak-Link Considerations and Lumped Element Model}

The spatial extent of the superfluid weak link measures a few coherence lengths on either side of the link. For a cell with dimensions much larger than $\xi$, the weak link is spatially separated from the other elements of the device. It can be treated as an independent fluidic element. For plate displacements that are much smaller than $\xi$, the superfluid condensate in the cell is robust and the plate is spatially separated from and independent of the weak link. Making the analogy to electrical circuits, we refer to this model as the lumped-element model and evaluate the total energy in the device as the sum of the energies stored in the independent elements.  

\subsection*{Circulation}
\begin{figure*}
\includegraphics[width=0.9\textwidth]{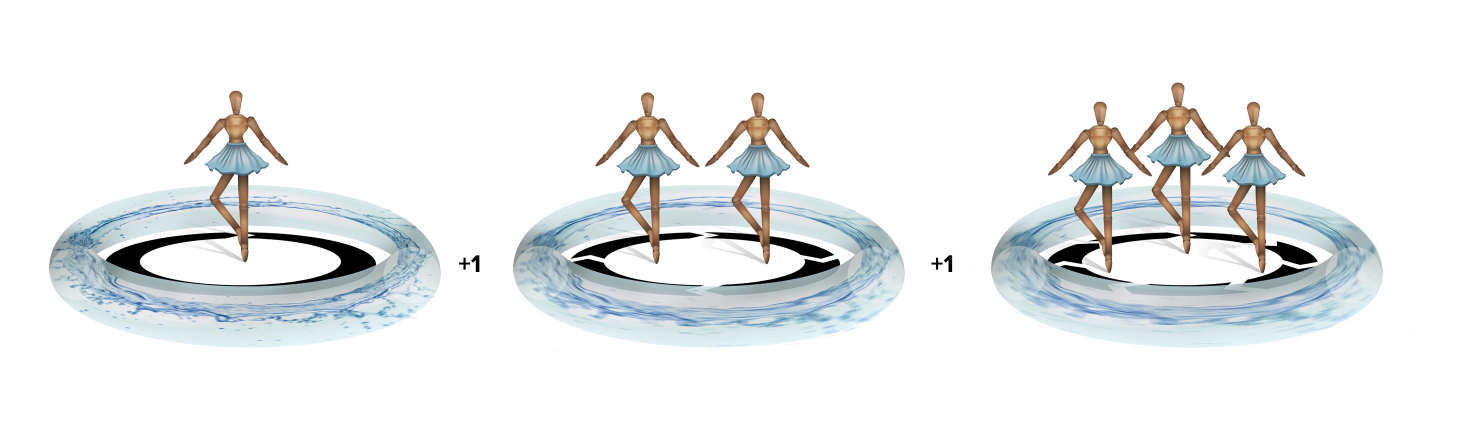}
    \caption{{{Artist's illustration of quantization of circulation. Circulating superfluid mass currents in the rings above have quantized ``circulation'' flux. The flux quantum is illustrated by a ballerina. As the supercurrent increases (left to right above), the circulation increases in discrete quanta. }} }
     \label{fig:circulation}
\end{figure*}

In Results, we define the generalized circulation $\mathcal{K}$ analogous to the generalized flux in the superconducting circuit case. In this context, we review briefly the concept of ``circulation'', commonly used in the physics of superfluids; we refer the reader to \cite{VW} for a more detailed description.  Consider a closed curve $\mathcal{C}$ confined in the superfluid condensate. The ``circulation'' $\kappa$ is defined as the line integral along $\mathcal{C}$,
\begin{equation}
    \kappa \equiv \ointop_{\mathcal{C}} \vec{v}\cdot d\vec{l}\,\,\,,
\end{equation}
where $\vec{v}$ is the local fluid velocity. For a simply connected container filled with superfluid, the circulation is identically zero,  $\kappa = 0$, since the line integral along any closed curve in the superfluid is given 
by the surface integral of the curl of the vector field $\vec{v}$, which is irrotational i.e.,  $\ointop_{\mathcal{C}} \vec{v}\cdot d\vec{l} = \int(\nabla\times\vec{v})\cdot d\vec{S} = 0$.
For superflow in an annular container, this line integral is not identically zero and can be finite valued. The superflow is given by the gradient of the superfluid phase, $\vec{v} = \frac{\hbar}{2m}\nabla\varphi$. Since $\varphi$ can only change in multiples of $2\pi$ at every given point on $\mathcal{C}$,
\begin{equation}
\label{circulationquantum}
    \kappa = \frac{\hbar}{2m}\ointop_{\mathcal{C}}\nabla\varphi\cdot d\vec{l} \equiv n\,\kappa_0\,\,\,\,;\,\,\,\,n = 0, \pm 1, \pm 2, ...,
\end{equation}
and the circulation $\kappa$ is quantized. 
It follows that a superfluid in an annulus carries quantized (persistent) currents. To help visualise this concept, an artist's illustration of circulation flux quantization is shown in Fig. \ref{fig:circulation}.

\subsection*{Analogue to Gate Voltage}
We discuss here the superfluid analogue of applying a gate voltage as the application of a constant external pressure to the superfluid in the SHOQ device. This could be a useful scheme to engineer SHOQ  device operation.

Consider a constant external pressure difference maintained between the two sides of the plate element in the SHOQ device, disregarding the weak link for the moment i.e., the fluid enclosed in the cell is at a hydrostatic pressure $P_{in}$ that is different from the pressure of the fluid outside $P_{out} \neq P_{in}$. This could be implemented by replacing the rigid bottom surface of the SHOQ cell by a stiff movable plate; with the plate motion controlled to apply a constant (additional) external pressure to the superfluid in the cell.   
In this case, the displacement of the plate in the simple harmonic regime is given by
\begin{equation}
\frac{kx}{A} = \delta P + P_0\,\,\,;\,\,\,P_0 = P_{in} - P_{out}\,\,\,.
\end{equation}
Here, $P_0$ is the constant pressure difference between the two sides of the plate and $\delta P$ is the pressure difference attributed to Josephson tunneling through the weak link. We assume sufficiently low pressure bias, $P_0$ such that $\delta\mu \ll {\bf{\Delta}}$, where $\bf{\Delta}$ is the superfluid gap, so that the superfluid remains in the hydrodynamic regime.
Following equations (\ref{phidot}-\ref{phidotdot}) for this case, we get
\begin{equation}
    \dot{\varphi} = -\frac{2m\,\delta P}{\rho \hbar} = -\frac{2m}{\rho\hbar}\left(\frac{kx}{A} - P_0\right)\,\,\,.
\end{equation}
It follows that the capacitive energy stored in the plate is
\begin{equation}
    E_{P} = \frac{1}{2} k x^2 
    = \frac{\rho^2 A^2 }{2 k} (\dot{\mathcal{K}} - \frac{P_0}{\rho})^2\,\,\,,
\end{equation}
using equations (\ref{phidotkappadot}) and (\ref{phidotdot}). 
Following through SHOQ Circuit Theory in Results to derive the canonical momentum $Q_g$, as in equation (\ref{Qdefn}) for this case, we obtain
\begin{equation}
    Q_g = \frac{ \rho^2 A^2}{k} (\dot{\mathcal{K}} - \frac{P_0}{\rho}) \equiv 2 m(n - n_g)\,\,\,,
\end{equation}
where we define  
\begin{equation}
    \label{ng}
    n_g \equiv 
    \frac{\rho A^2 P_0}{2 k m}\,\,\,.
\end{equation}  
The resulting Hamiltonian is
\begin{equation}
    \label{GatedH}
    \mathcal{H}_{SHOQ} =  E_C (n - n_g)^2 - E_J \cos(2\pi\frac{\mathcal{K}}{\kappa_0})\,\,\,,
\end{equation}
with definitions of $E_C$ and $E_J$ as in equation (\ref{ECJanalogues}). We thus show that applying a constant pressure  to the superfluid in the  SHOQ device is analogous to gating by a constant voltage in the transmon circuit, 
{in the context of the device Hamiltonian.}

\subsection*{SHOQ Coupling to a Superconducting Qubit}
\begin{figure}
    \centering
    \includegraphics[width=0.8\linewidth]{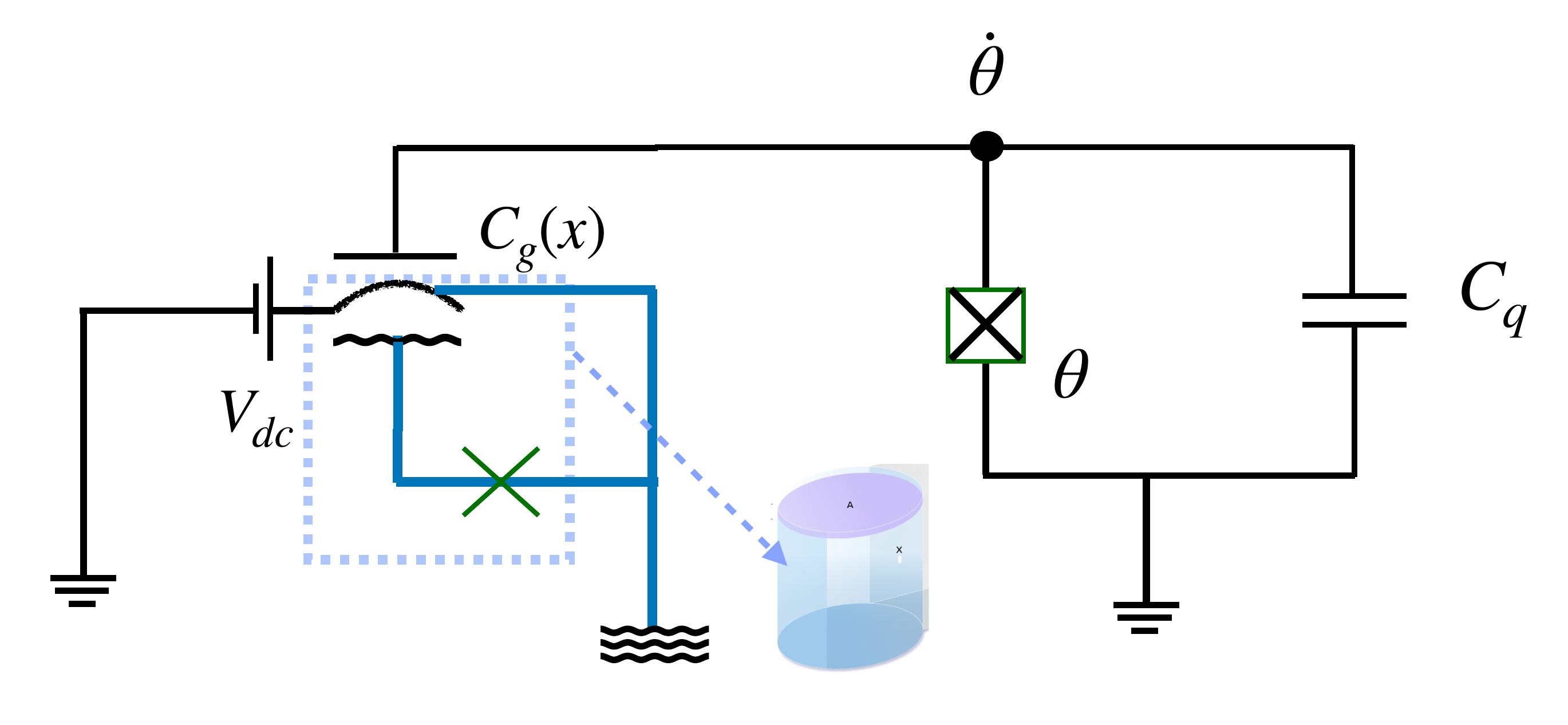}
    \caption{Circuit diagram for coupling between a transmon qubit and the SHOQ device. The capacitance  $C_g(x)$ is discussed in the text. }
    \label{fig:SHOQ-SCqubitCoupling}
\end{figure}
{We work out the coupling for a  circuit comprising a transmon qubit and a SHOQ device, see Fig. \ref{fig:SHOQ-SCqubitCoupling}. 
The capacitance $C_g(x)$ is a function of the displacement, $x$ of the elastic plate component of the SHOQ device. For small displacements $x$ (in particular, $x \ll d$ where $d$ is the equilibrium separation between the capacitor plates corresponding to $x=0$), $C_g(x)$ may be approximated by its Taylor expansion to linear order:
\begin{equation}
\label{Cexp}
C_g(x) = C_g(0) + C_g'(0)\,x + \mathcal{O}(x^2) 
\,\,\,.
\end{equation}
The resulting circuit Lagrangian is
\begin{eqnarray}
\label{couplingLagrangian}
    \nonumber
    \mathcal{L} &=& \frac{\hbar^2}{8 e^2}C_q \dot{\theta}^2 + \mathcal{E}_J\cos{\theta}  
    + \frac{\rho^2 \hbar^2 A^2}{8 k m^2}\dot{\varphi}^2 + E_J\cos{\varphi}\\
    &+& \frac{1}{2} (C_g(0) + C'_g(0)\frac{\rho\hbar A}{2mk}\dot\varphi)(V_{dc}-\frac{\hbar}{2e}\dot{\theta})^2\,\,\,.
\end{eqnarray}
Here, $\theta$ is the phase difference across the superconducting Josephson junction, $\mathcal{E}_J$ the Josephson energy, and $V_{dc}$ a constant DC voltage bias applied to the elastic plate component of the SHOQ device. The mechanical motion of the plate gives rise to a ``motional gate charge" that couples the SHOQ degree of freedom $\dot\varphi$ to the SC qubit degree of freedom $\dot\theta$. }
{We set $\dot\theta/V_{dc} \ll 1$
(while ensuring $V_{dc}$ is not too large to affect the coherence time of the SC qubit). 
We consider the first term in this perturbative expansion as the coupling between the SHOQ and SC degrees of freedom, $\dot\varphi$ and $\dot\theta$, respectively. 
In this regime,
\begin{equation}
\label{largebiasregime}
    \frac{1}{2}C_g(x)(V_{dc}-\frac{\hbar}{2e}\dot{\theta})^2 \sim \frac{1}{2}C_g V_{dc}^2 (1 -\frac{\hbar}{e}\frac{\dot{\theta}}{V_{dc}})\,\,\,.
\end{equation}
The Lagrangian in equation(\ref{couplingLagrangian}) may now be approximated as
\begin{eqnarray}
     \mathcal{L} &=& \frac{\hbar^2}{8 e^2}C_t \dot{\theta}^2 + \mathcal{E}_J\cos{\theta} + \frac{ \hbar^2 }{16 E_C}\dot{\varphi}^2 + E_J\cos{\varphi}\\
     \nonumber
     &-& C_{g}(0)V_{dc}\frac{\hbar}{2e}\dot{\theta} - C'_g V_{dc}^2\frac{\hbar}{\sqrt{8kE_C}}\dot{\varphi} + C'_g V_{dc}\frac{\hbar^2\dot{\theta}\dot{\varphi}}{2e\sqrt{8kE_C}}\\
     \nonumber
     &+& \frac{C_{g}(0)}{2} V_{dc}^2\,\,\,.
\end{eqnarray}
Here we have used equation(\ref{ECJanalogues}) and $C_t \equiv C_q + C_g(0)$.
Using the values for electrical parameters for the mechanical resonator used in \cite{HybridCouplingNature:2013},where $\omega_m = 72$ MHz and $x_{zp} = 4$ fm, 
$C'_g = 26$ nF/m, $V_{dc} = 1$ V and $C_q = 61$ fF used in Ref[35], we find 
\begin{equation}
    \frac{{C'_g}^2 V_{dc}^2}{k_{SHOQ} C_q} 
    \ll 1\,\,\,. 
\end{equation}
We set $ 1 - \frac{{C'_g}^2 V_{dc}^2}{k C_q} \sim 1$ henceforth.
Defining conjugate variables
\begin{equation}
    \Pi_1 = \frac{\partial \mathcal{L}}{\partial \dot\theta}\,\,\,;\,\,\,\Pi_2 = \frac{\partial \mathcal{L}}{\partial \dot\varphi}\,\,\,,
\end{equation}
we now get a pair of linear equations in $\dot\theta$ and $\dot\varphi$ with solutions,
\begin{eqnarray}
\nonumber
    \dot\theta &=& \frac{4e^2}{\hbar^2 C_t}\Pi_1 - \Lambda\Pi_2 + \Big(\frac{2C_{g}(0) V_{dc} e}{\hbar C_t} - \frac{2{C'_g}^2 V_{dc}^3 e}{\hbar k C_q}\Big)\\
    \dot\varphi &=& \frac{8 E_C}{\hbar^2}\Pi_2 - \Lambda\Pi_1 +\Lambda \frac{C_t V_{dc} \hbar}{2e}\,\,\,,
\end{eqnarray}
where $\Lambda \equiv \frac{2C'_g V_{dc} e}{
\hbar^2 
C_t}\sqrt{\frac{8 E_C}{k}}$.
Now, the Hamiltonian $\mathcal{H}$ is given by
\begin{equation}
\label{TransformtoHamiltonian}
    \mathcal{H} = \Pi_1\dot\theta + \Pi_2\dot\varphi - \mathcal{L}\,\,\,.
\end{equation}
Identifying $\Pi_1 = \frac{\hbar^2 C_t}{e^2} \dot\theta = \hbar n_q$ and $\Pi_2 = \frac{\kappa_0}{2\pi} 2m n = \hbar n$, where $n_q$ and $n$ are the number operators for the SC and SHOQ quanta, respectively, the coupling term $\mathcal{H}_{q,SHOQ}$ is given by the terms in $\mathcal{H}$ that are  $\propto \Pi_1\Pi_2$,
\begin{equation}
    \mathcal{H}_{q,SHOQ} = -\Pi_1\Pi_2 \Big(-\Lambda - \Lambda^2\frac{\hbar^4C_q}{2e^2 E_C}\Big)\,\,\,.
\end{equation}
We use the electrical parameters reported in \cite{HybridCouplingNature:2013} and use values for $C_g(0)$ set by typical piezo-sensing capacitive elements (used in the acoustic sensing context) in \cite{SourisDavis:2017} to  set $C_t = C_q\times 0.17$ nF, and find the magnitude of $\hbar\Lambda$ to be $\sim 1.8$ MHz. 
The regime of operation defined in equation (\ref{largebiasregime}) is consistent with the electrical parameters that apply to the coupling of a SC qubit to a mechanical resonator. 
Using the values of $V_{dc}$, $C_q$ and the root mean squared value of the number operator for the SC qubit 
we find the perturbative regime holds i.e., $n_q \sim \mathcal{O}(1)$, $\frac{\hbar}{2e}\dot{\theta} \sim \frac{\hbar}{2e}\frac{2e} {C_q} n_q  \ll V_{dc}$.}
{The calculation above is not a rigorous calculation from a full Lagrangian for the circuit shown in Fig.\ref{fig:SHOQ-SCqubitCoupling}. However, within the experimental parameters applicable to the analogous mechanical scheme 
we find a tractable solution to estimate the strength of coupling of the SHOQ device to a SC qubit. We  note that a complete analysis of the additional physics that enters when $V_{dc}$ is outside the parameter range 
is also beyond the scope of this calculation.}

\subsection*{Sources of Dissipation}

We discuss here estimates for the dissipation rates arising from viscous damping, the mechanocaloric effect, damping due to two-level systems in the substrate, single quasiparticle tunneling and  weak-link dissipation in the SHOQ device.
 A model detailing these sources of dissipation has been developed for superfluid Helmholtz resonators \cite{SourisDavis:2017}. We apply this model to obtain an estimate of the SHOQ device's quality factor, $Q$ given by the inverse of the total dissipation rate from the sources discussed below.

Among the sources of dissipation discussed in \cite{SourisDavis:2017} the dominant source at low temperatures is dissipation from single quasiparticle tunneling through the superfluid weak link. The details of the mechanism for this tunneling are specific to the exact form of the superfluid order parameter on both sides of the weak link. Therefore, this source of dissipation is inherently tied to textural dissipative effects across the weak link. For a uniform homogeneous B-phase texture across the weak link as assumed in our SHOQ device design, we make an  estimate for dissipation arising from single quasiparticle tunneling using the theory developed for qubits using Josephson links with BCS superconductors \cite{Glazman:2011}.  
We find the dissipation rate from single quasiparticle tunneling,  
$     Q_{sqt}^{-1} =
  \frac{1}{\pi} \sqrt{\frac{2\pi k_B T}{\hbar\omega_p}}\,e^{-\frac{{\bf{\Delta}}}{k_B T}}$
   . 
We estimate the relaxation time due to this source, $T_{sqt} = Q_{sqt}/\omega_p$ for various SHOQ design in Table \ref{tab:SHOQDesigns}.

At finite temperature, there exist both normal and superfluid components in the SHOQ device. The normal component of the liquid is viscous and remains clamped to the cell. 
Viscous damping of the motion of the plate is a source of dissipation. This is directly proportional to the normal-fluid density and vanishes in the $T=0$ limit. Since the proportion of normal component is negligible at the operating temperature of $\sim 0.4\, mK$ of the SHOQ device,  this form of dissipation is vanishingly small compared to $Q_{sqt}^{-1}$.

When normal fluid is clamped to the oscillating plate and superfluid moves in the cell, a temperature difference is driven between the cell and the reservoir referred to as the mechanocaloric effect. 
To evaluate the magnitude of this effect, we use the entropy density, fluid compressibility, Kapitza resistance and specific heats of $^3$He in the model developed for $^4$He in \cite{SourisDavis:2017}. We find the dissipation rate from the mechanocaloric effect for the $^3$He case to be much smaller than that for the $^4$He cell at temperatures below 1 K. The latter is estimated to be $\lesssim 10^{-6}$, much smaller than $Q_{sqt}^{-1}$.
Further, weak link experiments \cite{RMPJosephson:2002} report that temperature variations are negligibly small in similar setups providing further justification to ignore this source of dissipation.

Two-level systems in the substrate material are a source of  damping and the detailed model for dissipation arising from this source has been analysed in cells of glass/quartz
\cite{SourisDavis:2017}. In the SHOQ case, the choice of material for the elastic plate affects this source of dissipation. For example, it is known that this dissipation is much less for quartz vs.\ glass  \cite{SourisDavis:2017}. The dissipation rate from this source depends largely on the geometry and is $\propto 1/A^2$. For $mm$-sized cells used in \cite{SourisDavis:2017}, the dissipation rate from substrate two-level systems is $\lesssim 10^{-5}$. With the SHOQ device being designed with much smaller plates than used in , we argue that the dissipation rate from this source is $\ll Q_{sqt}^{-1}$.

\section*{Acknowledgments}

We are grateful for fruitful discussions with William Halperin, John Davis, Elena Lupo, Max Cykiert and Alexander Shook. 
This research was funded in whole by the Daphne Jackson Trust UK, the Engineering and Physical Sciences Research Council UK and in part by the Quantum For Science programme of the Science and Technologies Facilities Council, UK via the International Science Partnership Fund grant no.~(ST/Y00518X/1). 
P.S. and E.G. conceived and designed the analysis. P.S. performed the calculations/analysis. J.K. assisted with the Coupling analysis. All authors equally wrote and edited the manuscript. The authors declare no competing interests. All data needed to evaluate the conclusions in the paper are present in the paper.

\section*{Competing Interests}
Patent filed : Patent Application Number GB2408827.0, Inventors : P. Sharma, E. Ginossar (University of Surrey). Current status : International filed (PCT). 

\section*{Data Availability Statement}
This manuscript does not report data generation or analysis.

\section*{Code Availability Statement}
This manuscript does not report any outputs generated from code.

\bibliography{library}

@Article{RMPJosephson:2002,
  author  = {Davis, J. C. and Packard, R. E},
  title   = {{\it{Superfluid  $^3{H}e$  {J}osephson {W}eak {L}inks}}},
  pages   = {741},
  url     = {},
  volume  = {74},
  journal = {Rev. Mod Phys.},
  year    = {2002},
}

@book{VW,
    author = {Vollhardt, Dieter and W\"{o}lfle, Peter},
    title = {{\it{The {S}uperfluid {P}hases of {H}elium 3}}},
    publisher = {Taylor \& Francis} ,
    year = {1990}
}

@unpublished{DavisEmail,
    author = {Shook, Alexander J. and Varga, Emil and Davis, John P.},
    title = {},
    note = {private communication}
}

@Article{SourisDavis:2017,
  author  = { Souris, Fabien and  Rojas, Xavier and Kim, Paul H. and Davis,John P.},
  title   = {{\it{Ultra-{L}ow {D}issipation {S}uperfluid {M}icromechanical {R}esonator}}},
  pages   = {044008},
  url     = {},
  volume  = {7},
  journal = {Phys. Rev. Applied},
  year    = {2017},
}

@Article{Glazman:2011,
    author = {Catelani, G. and Koch, J. .and Frunzio, L. and Schoelkopf, R. J. and Devoret, M. H. and Glazman, L. I.},
    title = {{\it{Quasiparticle {R}elaxation of {S}uperconducting {Q}ubits in the {P}resence of {F}lux}}},
    pages = {077002},
    volume = {106},
    journal = {Phy. Rev. Lett.},
    year = {2011},
}

@article{BW,
  title = {{\it{Superconductivity with {P}airs in a {R}elative $p$-Wave}}},
  author = {Balian, R. and Werthamer, N. R.},
  journal = {Phys. Rev.},
  volume = {131},
  issue = {4},
  pages = {1553--1564},
  numpages = {0},
  year = {1963},
  month = {Aug},
  publisher = {American Physical Society},
  doi = {10.1103/PhysRev.131.1553},
  url = {https://link.aps.org/doi/10.1103/PhysRev.131.1553}
}

@article{Hoskinson:2005,
  author = {Hoskinson, E. and Packard, R. and Haard, T.},
  title = {{\it{Quantum {W}histling in {S}uperfluid {H}elium-4}}},
  journal = {Nature.},
  volume = {433},
  issue = {},
  pages = {376},
  numpages = {0},
  year = {2005},
  month = {},
  publisher = {Springer},
  doi = { https://doi.org/10.1038/433376a},
  url = {https://www.nature.com/articles/433376a#Abs13}
}

@Inbook{Sato2019,
author={Sato, Y.
and Hoskinson, E.
and Packard, R. E.},
editor={Tafuri, Francesco},
title={{\it{Josephson {E}ffects in {S}uperfluid {H}elium : {F}undamentals and {F}rontiers of the {J}osephson {E}ffect}}},
year={2019},
publisher={Springer International Publishing},
address={Cham},
pages={765--810},
isbn={978-3-030-20726-7},
doi={10.1007/978-3-030-20726-7_19},
url={https://doi.org/10.1007/978-3-030-20726-7_19}
}

@article{ThunebergReview:2005,
    author = {Thuneberg, Erkki},
    title = {{\it{Theory of {J}osephson {P}henomena in {S}uperfluid  $^3${H}e}}},
    journal = {AIP Conference Proceedings},
    volume = {850},
    number = {1},
    pages = {103-108},
    year = {2006},
    month = {09},
    issn = {0094-243X},
    doi = {10.1063/1.2354625},
    url = {https://doi.org/10.1063/1.2354625},
}

@article{WendinReview:2017,
    author = {Wendin, G},
    title = {{\it{Quantum {I}nformation {P}rocessing with
{S}uperconducting {C}ircuits: {A} {R}eview"}}},
    journal = {Rep. Prog. Phys.},
    volume = {80},
    pages = {106001},
    year = {2017},
    doi = {10.1088/1361-6633/aa7e1a}
}

@Inbook{Devoret:1997,
author={Devoret, M. H.},
editor={Reynaud, S. and Giacobino, E. and
Zinn-Justin, J.},
title={{\it{Quantum {F}luctuations in {E}lectrical {C}ircuits}}},
bookTitle={Quantum Fluctuations (Les Houches
Session LXIII)},
year={1997},
publisher={Elsevier},
address={},
pages={351–386},
isbn={},
doi={},
url={}
}

@article{RMPWheatley:1975,
    author = {Wheatley, John C.},
    title = {{\it{Experimental {P}roperties of {S}uperfluid  $^3${H}e}}},
    journal = {Rev. Mod. Phys.},
    volume = {47},
    pages = {415},
    year = {1975},
    doi = {}
}

@Article{Schwab:2017,
    title = {{\it{Ultra-{H}igh {Q} {A}coustic {R}esonance in {S}uperfluid  $^4${H}e}}},
    author = {De Lorenzo, L. A. and Schwab, K. C.},
    journal = {J. Low Temp. Phys.},
    volume = {1876},
    pages = {233},
    year = {2017}, 
    doi = {}
}

@article{PRLMillisecQubit:2023,
  title = {{\it{Millisecond {C}oherence in a {S}uperconducting {Q}ubit}}},
  author = {Somoroff, Aaron and Ficheux, Quentin and Mencia, Raymond A. and Xiong, Haonan and Kuzmin, Roman and Manucharyan, Vladimir E.},
  journal = {Phys. Rev. Lett.},
  volume = {130},
  issue = {26},
  pages = {267001},
  numpages = {6},
  year = {2023},
  month = {Jun},
  publisher = {American Physical Society},
  doi = {10.1103/PhysRevLett.130.267001},
  url = {https://link.aps.org/doi/10.1103/PhysRevLett.130.267001}
}

@article{NatureMillisecTransmon:2021,
    author = {Place, A.P.M. and Rodgers, L.V.H.  and Mundada, P. and others},
    title = {{\it{New {M}aterial {P}latform for {S}uperconducting {T}ransmon {Q}ubits with {C}oherence {T}imes {E}xceeding 0.3 milliseconds}}},
    journal = {Nat. Commun.},
    volume = {12},
    pages = {1779},
    year = {2021},
    doi = {}
}

@article{HybridCouplingNature:2013,
    author = {Pirkkalainen, J.M. and Cho, S. U. and Li, Jian and Paraoanu, G. S. and Hakonen, P. J. and Sillanp{\"{a}}{\"{a}}, M. A.},
    title = {{\it{Hybrid {C}ircuit {C}avity {Q}uantum {E}lectrodynamics with a {M}icromechanical {R}esonator"}}},
    journal = {Nature},
    volume = {494},
    pages = {211},
    year = {2013},
    doi = {}
}

@article{QUEST-DMC:2024,
    author = {Autti, S. and Casey, A. and Eng, N. and others} ,
    title = {{\it{{QUEST-DMC}: {B}ackground {M}odelling and {R}esulting {H}eat {D}eposit for a {S}uperfluid {H}elium-3 {B}olometer}}},
    journal = {J Low Temp Phys},
    volume = {215},
    year = {2024},
    pages = {465},
    doi = {https://doi.org/10.1007/s10909-024-03142-w}
}

@article{SaundersAB:2024,
    author = {Heikkinen, P.J. and Eng, N. and Levitin, L.V. and others},
    title = {{\it{Nanofluidic {P}latform for {S}tudying the {F}irst-{O}rder {P}hase {T}ransitions in {S}uperfluid {H}elium-3}}} ,
    journal = {J Low Temp Phys},
    year = {2024},
    volume = {215},
    pages = {477}
}

@article{ShookPRL:2024,
    author = {Shook, Alexander J. and  Varga, Emil and  Boettcher, Igor and Davis, John P. } ,
    title = {{\it{Surface {S}tate {D}issipation in {C}onfined  $^3${H}e-{A}}}} ,
    journal = {Phys. Rev. Lett.},
    volume = {132}, 
    pages = {156001},
    year = {2024}
}

@article{VorontsovSauls:2007,
    title = {{\it{Crystalline {O}rder in {S}uperfluid  {F}ilms}}},
    author = {Vorontsov, A. B. and  Sauls, J. A.},
    journal = {Phys. Rev. Lett.},
    volume = {98}, 
    pages = {045301},
    year = {2007}
}

@article{Levitin:2019,
    title = {{\it{Evidence for a {S}patially {M}odulated {S}uperfluid {P}hase of $^3{H}e$ under {C}onfinement}}},
    author = {Levitin, Lev V. and Yager,  Ben and others},
    journal = {Phys. Rev. Lett.},
    volume = {122}, 
    pages = {085301},
    year = {2019}}

@article{DavisPDW:2020,
    title = {{\it{Stabilized {P}air {D}ensity {W}ave via {N}anoscale {C}onfinement of {S}uperfluid  $^3{H}e$}}},
    author = {Shook, A. J. and Vadakkumbatt, V. and   Yapa, P. Senarath and others},
    journal = {Phys. Rev. Lett.},
    volume = {124}, 
    pages = {015301},
    year = {2020}
}

@article{AuttiHQV:2016,
    title = {{\it{Observation of {H}alf-{Q}uantum {V}ortices in {T}opological {S}uperfluid   $^3{H}e$}}},
    author = {Autti, S. and  Dmitriev, V. V. and Mäkinen, J. T. and others},
    journal = {Phys. Rev. Lett.},
    volume = {117},
    pages = {255301 },
    year = {2016}
}

@article{ClelandSET:2010,
    title = {{\it{Quantum {G}round {S}tate and {S}ingle-{P}honon
{C}ontrol of a {M}echanical {R}esonator}}},
    author = {O’Connell, A. D. and Hofheinz, M. and Ansmann, M. and others},
    journal = {Nature},
    volume = {464},
    pages = {697},
    year = {2010}
}

@article{Teufel:2008,
    title = {{\it{Measuring {N}anomechanical {M}otion with a
{M}icrowave {C}avity {I}nterferometer}}},
    author = {Regal, C. A. and Teufel, J. D. and Lehnert, K. W.},
    journal = {Nat. Phys.},
    volume = {4},
    pages = {555},
    year = {2008}
}

@article{Schwab:2003,
    title = {{\it{Quantum {M}easurement of a {C}oupled {N}anomechanical {R}esonator–{C}ooper-pair {B}ox {S}ystem}}},
    author = { Irish, Twyeffort E. K. and Schwab, K.},
    journal = {Phys. Rev. B},
    volume = {68}, 
    pages = {155311},
    year = {2003}
}

@article{Harris:2008,
    author = {Jayich, A. M. and Sankey, J. C. and Zwickl, B. M. and others},
title = {{\it{Dispersive {O}ptomechanics: {A} {M}embrane
inside a {C}avity}}},
    journal = {New Journal of Phys.},
    volume = {10},
    pages = {095008},
    year = {2008}
}

@article{Kawakami:2024,
    title = {{\it{Quantum {C}omputing using {F}loating {E}lectrons on {C}ryogenic {S}ubstrates: {P}otential and {C}hallenges}}},
    author = {Jennings, A. and Zhou, X. and Grytsenko, I. and Kawakami, E.},
    journal = {Appl. Phys. Lett.},
    volume = {124}, 
    pages = {120501},
    year = {2024} 
}

@article{Pollanen:2024,
    title = {{\it{Coulomb {I}nteraction-{D}riven {E}ntanglement of {E}lectrons on {H}elium}}},
    author = {Beysengulov, Niyaz R. and Schøyen, O. S. and  Bilek, Stian D. and others},
    journal = {PRX Quantum},
    volume = {5},
    pages = {030324} ,
    year = {2024}
}

@article{Tomography:2013,
    author = {Kirchmair, G.  and Vlastakis, B. and Leghtas, Z.  and others}, 
    title = {{\it{Observation of {Q}uantum {S}tate {C}ollapse and {R}evival due to the {S}ingle-{P}hoton {K}err {E}ffect }}},
    journal = {Nature},
    volume = {495}, 
    pages = {205},
    year = {2013},
     doi = {https://doi.org/10.1038/nature11902}
}

@book{VolovikBook,
    author = {Volovik, Grigory E.}, 
    title = {{\it{The Universe in a Helium Droplet}}},
    publisher = {International Series of Monographs on Physics (Oxford, 2009; online edn, Oxford Academic, 1 Jan. 2010)},
    doi = {https://doi.org/10.1093/acprof:oso/9780199564842.001.0001}
}

@article{SaulsJosephson:1990,
    author = {Thuneberg, E. V. and Kurkij\"{a}rvi, J. and Sauls, J. A.},
    title = {{\it{Quasiclassical {T}heory of the {J}osephson {E}ffect in {S}uperfluid  $^3${H}e}}},
    journal = {Physica B},
    volume = {165-166},
    pages = {755},
    year = {1990}
}

@article{ViljasThuneberg:2004,
    author = {Viljas, J. K. and Thuneberg, E. V.} ,
    title = {{\it{Textural {E}ffects and {S}pin-{W}ave {R}adiation in {S}uperfluid  $^3${H}e {W}eak {L}inks}}},
    journal = {J. Low Temp. Phys.},
    volume = {136},
    pages = {329},
    year = {2004}
}

@article{ViljasThunebergPRL,
  title = {{\it{Dissipative {C}urrents in {S}uperfluid $^{3}${H}e {W}eak {L}inks}}},
  author = {Viljas, J. K. and Thuneberg, E. V.},
  journal = {Phys. Rev. Lett.},
  volume = {93},
  issue = {20},
  pages = {205301},
  numpages = {4},
  year = {2004},
  month = {Nov},
  publisher = {American Physical Society},
  doi = {10.1103/PhysRevLett.93.205301},
  url = {https://link.aps.org/doi/10.1103/PhysRevLett.93.205301}
}

@book{Spindry,
    author = {Low Temperature Physics Group, Northwestern University},
    title = {$^{3}${H}e Calculator},
    publisher = {https://spindry.phys.northwestern.edu/he3.htm},
    year = {}
}

@article{
MechanicalQubit2024,
author = {Yu Yang  and Igor Kladarić  and Maxwell Drimmer  and Uwe von Lüpke  and Daan Lenterman  and Joost Bus  and Stefano Marti  and Matteo Fadel  and Yiwen Chu },
title = {{\it{A Mechanical Qubit}}},
journal = {Science},
volume = {386},
number = {6723},
pages = {783-788},
year = {2024},
doi = {},
URL = {https://www.science.org/doi/abs/10.1126/science.adr2464},
eprint = {},
}

@article{Sato2012,
    author = {Y Sato and R E Packard},
    title = {{\it{Superfluid Helium Quantum Interference Devices: Physics and 
             Applications}}},
    journal = {Rep. Prog. Phys.},
    volume = {75},
    year = {2012},
    pages = {016401} ,
    doi = {doi:10.1088/0034-4885/75/1/016401}
}

@Article{slussarenko2019photonic,
  author    = {Slussarenko, Sergei and Pryde, Geoff J},
  journal   = {Appl. Phys. Rev.},
  title     = {{\it{Photonic Quantum Information Processing: A Concise Review}}},
  year      = {2019},
  number    = {4},
  volume    = {6},
  pages = {041303},
  journal  = {Applied Physics Reviews},
  publisher = {AIP Publishing},
}

@Article{O_Brien_2009,
  author    = {O’Brien, Jeremy L. and Furusawa, Akira and Vučković, Jelena},
  journal   = {Nat. Photonics},
  title     = {{\it{Photonic Quantum Technologies}}},
  year      = {2009},
  issn      = {1749-4893},
  month     = dec,
  number    = {12},
  pages     = {687},
  volume    = {3},
  doi       = {10.1038/nphoton.2009.229},
  fjournal  = {Nature Photonics},
  publisher = {Springer Science and Business Media LLC},
}

@Article{Flamini_2018,
  author    = {Flamini, Fulvio and Spagnolo, Nicolò and Sciarrino, Fabio},
  journal   = {Rep. Prog. Phys.},
  title     = {{\it{Photonic Quantum Information Processing: A Review}}},
  year      = {2018},
  issn      = {1361-6633},
  month     = nov,
  number    = {1},
  pages     = {016001},
  volume    = {82},
  doi       = {10.1088/1361-6633/aad5b2},
  fjournal  = {Reports on Progress in Physics},
  publisher = {IOP Publishing},
}

@Article{Bruzewicz_2019,
  author    = {Bruzewicz, Colin D. and Chiaverini, John and McConnell, Robert and Sage, Jeremy M.},
  journal   = {Appl. Phys. Rev.},
  title     = {{\it{Trapped-ion Quantum Computing: Progress and Challenges}}},
  year      = {2019},
  issn      = {1931-9401},
  month     = {05},
  number    = {2},
  volume    = {6},
  pages = {021314},
  doi       = {},
  fjournal  = {Applied Physics Reviews},
  publisher = {AIP Publishing},
}

@Article{HAFFNER_2008,
  author    = {Haffner, H and Roos, C and Blatt, R},
  journal   = {Phys. Rep.},
  title     = {{\it{Quantum Computing with Trapped Ions}}},
  year      = {2008},
  issn      = {0370-1573},
  month     = dec,
  number    = {4},
  pages     = {155},
  volume    = {469},
  doi       = {10.1016/j.physrep.2008.09.003},
  fjournal  = {Physics Reports},
  publisher = {Elsevier BV},
}

@article{monroe2013scaling,
  title={{\it{Scaling the ion trap quantum processor}}},
  author={Monroe, Christopher and Kim, Jungsang},
  journal={Science},
  volume={339},
  number={6124},
  pages={1164--1169},
  year={2013},
  publisher={American Association for the Advancement of Science}
}

@article{henriet2020quantum,
  title={{\it{Quantum Computing with Neutral Atoms}}},
  author={Henriet, Lo{\"\i}c and Beguin, Lucas and Signoles, Adrien and Lahaye, Thierry and Browaeys, Antoine and Reymond, Georges-Olivier and Jurczak, Christophe},
  journal={Quantum},
  volume={4},
  pages={327},
  year={2020},
  publisher={Verein zur F{\"o}rderung des Open Access Publizierens in den Quantenwissenschaften}
}

@Article{saffman2016quantum,
  author    = {Saffman, Mark},
  journal   = {J. Phys. B: At. Mol. Opt. Phys.},
  title     = {{\it{Quantum Computing with Atomic Qubits and Rydberg Interactions: Progress and Challenges}}},
  year      = {2016},
  number    = {20},
  pages     = {202001},
  volume    = {49},
  fjournal  = {Journal of Physics B: Atomic, Molecular and Optical Physics},
  publisher = {IOP Publishing},
}

@article{Chatterjee_2021,
   title={{\it{Semiconductor Qubits in Practice}}},
   volume={3},
   ISSN={2522-5820},
   url={http://dx.doi.org/10.1038/s42254-021-00283-9},
   DOI={10.1038/s42254-021-00283-9},
   number={3},
   journal={Nature Reviews Physics},
   publisher={Springer Science and Business Media LLC},
   author={Chatterjee, Anasua and Stevenson, Paul and De Franceschi, Silvano and Morello, Andrea and de Leon, Nathalie P. and Kuemmeth, Ferdinand},
   year={2021},
   month=feb, pages={157} 
}

@Article{Burkard_2023,
  author    = {Burkard, Guido and Ladd, Thaddeus D. and Pan, Andrew and Nichol, John M. and Petta, Jason R.},
  journal   = {Rev. Mod. Phys.},
  title     = {{\it{Semiconductor Spin Qubits}}},
  year      = {2023},
  issn      = {1539-0756},
  month     = jun,
  number    = {2},
  volume    = {95},
  pages = {025003},
  doi       = {},
  fjournal  = {Reviews of Modern Physics},
  publisher = {American Physical Society (APS)},
}

@Article{Vandersypen2019a,
  author   = {Vandersypen, Lieven M K and Eriksson, Mark A},
  title    = {{\it{Quantum Computing with Semiconductor Spins}}},
  pages    = {38--45},
  volume   = {72},
  fjournal = {Physics Today},
  journal  = {Phys. Today},
  year     = {2019},
}

@Article{Bravyi_2022,
  author    = {Bravyi, Sergey and Dial, Oliver and Gambetta, Jay M. and Gil, Darío and Nazario, Zaira},
  journal   = {J. Appl. Phys.},
  title     = {{\it{The Future of Quantum Computing with Superconducting Qubits}}},
  year      = {2022},
  issn      = {1089-7550},
  month     = {10},
  number    = {16},
  volume    = {132},
  doi       = {},
  pages = {160902},
  fjournal  = {Journal of Applied Physics},
  publisher = {AIP Publishing},
}

@Article{Kjaergaard_2020,
  author    = {Kjaergaard, Morten and Schwartz, Mollie E. and Braumüller, Jochen and Krantz, Philip and Wang, Joel I.-J. and Gustavsson, Simon and Oliver, William D.},
  journal   = {Annu. Rev. Conden. Ma. P.},
  title     = {{\it{Superconducting Qubits: Current State of Play}}},
  year      = {2020},
  issn      = {1947-5462},
  month     = mar,
  number    = {1},
  pages     = {369},
  volume    = {11},
  doi       = {10.1146/annurev-conmatphys-031119-050605},
  fjournal  = {Annual Review of Condensed Matter Physics},
  publisher = {Annual Reviews},
}

@article{Zhang_2020,
  author    = {Zhang, Gang and Cheng, Yuan and Chou, Jyh-Pin
and   Gali, Adam},
  journal   = {Appl. Phys. Rev. },
  title     = {{\it{Material Platforms for Defect Qubits and Single-Photon Emitters}}},
  year      = {2020},
  issn      = {},
  month     = {},
  number    = {},
  pages     = {031308},
  volume    = {7},
  doi       = {https://doi.org/10.1063/5.0006075},
}

@article{PDWinHe3_2020,
  title = {{\it{Stabilized Pair Density Wave via Nanoscale Confinement of Superfluid $^{3}\mathrm{He}$}}},
  author = {Shook, A. J. and Vadakkumbatt, V. and Senarath Yapa, P. and Doolin, C. and Boyack, R. and Kim, P. H. and Popowich, G. G. and Souris, F. and Christani, H. and Maciejko, J. and Davis, J. P.},
  journal = {Phys. Rev. Lett.},
  volume = {124},
  issue = {1},
  pages = {015301},
  numpages = {6},
  year = {2020},
  month = {Jan},
  publisher = {American Physical Society},
  doi = {10.1103/PhysRevLett.124.015301},
  url = {https://link.aps.org/doi/10.1103/PhysRevLett.124.015301}
}

@article{SuperfluidOptomechanics_2021,
  title = {{\it{Superfluid Optomechanics With Phononic Nanostructures}}},
  author = {Spence, S. and Koong, Z.X. and Horsley, S.A.R. and Rojas, X.},
  journal = {Phys. Rev. Appl.},
  volume = {15},
  issue = {3},
  pages = {034090},
  numpages = {22},
  year = {2021},
  month = {Mar},
  publisher = {American Physical Society},
  doi = {10.1103/PhysRevApplied.15.034090},
  url = {https://link.aps.org/doi/10.1103/PhysRevApplied.15.034090}
}

@article{SuperfuidGWDetector_2021,
  title = {{\it{Prototype superfluid gravitational wave detector}}},
  author = {Vadakkumbatt, V. and Hirschel, M. and Manley, J. and Clark, T. J. and Singh, S. and Davis, J. P.},
  journal = {Phys. Rev. D},
  volume = {104},
  issue = {8},
  pages = {082001},
  numpages = {9},
  year = {2021},
  month = {Oct},
  publisher = {American Physical Society},
  doi = {10.1103/PhysRevD.104.082001},
  url = {https://link.aps.org/doi/10.1103/PhysRevD.104.082001}
}

@article{Koch:2007,
  title = {{\it{Charge-insensitive qubit design derived from the Cooper pair box}}},
  author = {Koch, Jens and Yu, Terri M. and Gambetta, Jay and Houck, A. A. and Schuster, D. I. and Majer, J. and Blais, Alexandre and Devoret, M. H. and Girvin, S. M. and Schoelkopf, R. J.},
  journal = {Phys. Rev. A},
  volume = {76},
  issue = {4},
  pages = {042319},
  numpages = {19},
  year = {2007},
  month = {Oct},
  publisher = {American Physical Society},
  doi = {10.1103/PhysRevA.76.042319},
  url = {https://link.aps.org/doi/10.1103/PhysRevA.76.042319}
}

@article{Whistling,
  author    = {Hoskinson, E. and Packard, R. and Haard, T.},
  journal   = {Nature },
  title     = {{\it{Quantum whistling in superfluid helium-4}}},
  year      = {2005},
  issn      = {},
  month     = {},
  number    = {},
  pages     = {376},
  volume    = {433},
  doi       = {https://doi.org/10.1038/433376a}
}
\end{document}